\documentclass[11pt]{article}
\pdfoutput=1
\usepackage{jheppub}
\usepackage[T1]{fontenc}
\usepackage{cancel,tensor,enumitem,fix-cm}
\usepackage{epsfig}
\usepackage{graphicx}
\usepackage[english]{babel}
\usepackage{amsmath,amssymb}
\usepackage{dsfont}
\usepackage{textcomp}
\usepackage{multirow}
\usepackage{booktabs}
\usepackage{overpic}
\usepackage{hyperref}
\usepackage{bm}
\usepackage{physics}
\usepackage[lofdepth,lotdepth]{subfig}
\usepackage{float}
\usepackage[dvipsnames]{xcolor}

\usepackage{tikz}
\usetikzlibrary{positioning}
\usetikzlibrary{intersections}
\usetikzlibrary{fadings} 
    \usetikzlibrary{arrows.meta} 
\usetikzlibrary{arrows}

\tikzfading[name=fade out,
inner color=transparent!0,
outer color=transparent!100]

\definecolor{cherryblossompink}{rgb}{1.0, 0.72, 0.77}
\definecolor{lightblue}{rgb}{0.68, 0.85, 0.9}

\usetikzlibrary{decorations.pathmorphing}
\usetikzlibrary{decorations.pathreplacing,decorations.markings}

\usetikzlibrary{backgrounds,automata}

\newcommand{\be}{\begin{equation}}
\newcommand{\ee}{\end{equation}}
\newcommand{\bea}{\begin{eqnarray}}
\newcommand{\eea}{\end{eqnarray}}

\newcommand{\beq}{\begin{equation}}
\newcommand{\eeq}{\end{equation}}

\def\({\left(}
\def\){\right)}
\def\[{\left[}
\def\]{\right]}

\subheader{\begin{flushright}
\texttt{IFT-UAM/CSIC-23-71}
\end{flushright}}

\title{Gravitation from optimized computation: Einstein and beyond}

\author[a]{Rafael Carrasco,}
\author[a]{Juan F. Pedraza,}
\author[b]{Andrew Svesko,}
\author[c]{and Zachary Weller-Davies}
\affiliation[a]{Instituto de F\'isica Te\'orica UAM/CSIC, Calle Nicol\'as Cabrera 13-15, Madrid 28049, Spain}
\affiliation[b]{Department of Physics and Astronomy, University College London, London WC1E 6BT, UK}
\affiliation[c]{Perimeter Institute for Theoretical Physics, Waterloo, ON N2L 2Y5, Canada}
\emailAdd{rafael.carrasco@ift.csic.es}
\emailAdd{j.pedraza@csic.es}
\emailAdd{a.svesko@ucl.ac.uk}
\emailAdd{zwellerdavies@pitp.ca}

\abstract{A new principle in quantum gravity, dubbed spacetime complexity, states that gravitational physics emerges
from spacetime seeking to optimize the computational cost of its quantum dynamics. Thus far, this principle has been realized at the linearized level, in holographic theories with Einstein gravity duals, assuming the so-called `Complexity-Volume' (CV) proposal. We expand on this proof in two significant directions. First, we derive higher-derivative gravitational equations by including appropriate corrections to the CV dictionary. Second, we show semi-classical equations arise by considering the leading bulk quantum corrections to CV. Our proof is valid for two-dimensional dilaton gravities, where the problem of semi-classical backreaction can be solved exactly. However, we argue the principle should hold more generally, leading us to a concrete proposal for bulk complexity of perturbative excited states in arbitrary dimensions. Our results demonstrate the robustness of spacetime complexity as a guiding principle to understand gravity in terms of quantum computation.}

\begin{document}
\maketitle
\flushbottom

\section{Introduction} \label{sec:intro}

The discovery that black holes carry a thermal entropy \cite{Bekenstein:1973ur,Hawking:1975vcx} leads to two fundamental insights: nature is holographic, and gravity is emergent. A concise realization of both features is captured by the Ryu-Takayanagi (RT) formula \cite{Ryu:2006bv}, a prescription to compute the entanglement entropy of subregions in quantum field theory states living on the asymptotic boundary of a `bulk' spacetime via the area of bulk codimension-2 surfaces. The RT formula is most precisely formulated in the $\text{AdS}/\text{CFT}$ correspondence, a specific instance of the holographic principle, where aspects of gravity in bulk asymptotically anti-de Sitter spacetime (AdS) are given a dual description in terms of a conformal field theory (CFT). In fact, the dynamics of spacetime itself emerges from the first law of entanglement \cite{Lashkari:2013koa,Faulkner:2013ica},
\beq \delta S_{A}=\delta\langle H_{A}\rangle\qquad \Longrightarrow \qquad \delta E_{\mu\nu}^{g}=0\;.\label{eq:firstlawent}\eeq
Here $S_{A}$ denotes the entanglement entropy of a CFT state $\rho_{A}$ confined to a boundary region $A$, $H_{A}\equiv e^{-\rho_{A}}$ is the modular Hamiltonian, and $\delta S_{A}$ is the change of entropy due to small perturbations to the state. Specifically, when $\rho_{A}=\rho_{A}^{(0)}+\delta\rho$, with $\rho_{A}^{(0)}$ being the vacuum state of a holographic CFT confined to a ball $A$, the first law of (holographic) entanglement 
is dual to the linearized gravitational equations of motion, $\delta E^{g}_{\mu\nu}=0$.\footnote{Non-linear corrections to gravitational equations of motion can also be obtained via a suitable generalization of the first law of entanglement \cite{Faulkner:2017tkh,Haehl:2017sot}. Prior work on the connection between gravitational dynamics and entanglement was given in \cite{Verlinde:2010hp,Verlinde:2016toy}. Moreover, in \cite{Jacobson:2015hqa} it was shown AdS/CFT need not be an input, and one is able to derive the full non-linear, semi-classical Einstein equations by assuming the vacuum of small causal diamonds is a maximal entropy state. This approach was extended in \cite{Bueno:2016gnv} where, however, only linearized equations of motion of higher-order theories are captured by the first law.} The latter observation may be interpreted as an information-theoretic version of  ``spacetime thermodynamics'' \cite{Jacobson:1995ab}, where the Einstein field equations are a consequence of assuming spacetime locally obeys the Clausius relation.\footnote{See \cite{Padmanabhan:2007en,Parikh:2009qs,Guedens:2011dy,Dey:2016zka,Parikh:2017aas,Parikh:2018anm,Svesko:2018qim,Alonso-Serrano:2022qvo} for routes to derive higher-curvature theories of gravity from equilibrium thermodynamics.} In the same spirit, (\ref{eq:firstlawent}) implies gravity emerges from ``spacetime entanglement''.

Entanglement, however, is not enough to describe all aspects of bulk gravitational physics \cite{Susskind:2014rva,Susskind:2014moa}. Specifically, the late time growth of the Einstein-Rosen bridge inside of eternal black holes is characterized by another dual information-theoretic quantity, namely, complexity. In an ordinary quantum mechanical setting, (computational) complexity refers to the smallest number of unitary operators, or gates, needed to obtain a particular target state from a given reference state, within a specified margin of error. Thus, complexity amounts to an optimization problem in computation. A precise definition of complexity in field theories remains an important area of investigation (cf. \cite{Chapman:2017rqy,Jefferson:2017sdb,Caputa:2017urj,Caputa:2017yrh,Chapman:2018hou,Hackl:2018ptj,Camargo:2019isp,Flory:2020eot,Flory:2020dja,Chagnet:2021uvi,Chandra:2022pgl}), nonetheless, one can ask what is the corresponding geometric dual of complexity of a holographic CFT. By now there are many conjectures \cite{Susskind:2014rva,Susskind:2014jwa,Stanford:2014jda,Couch:2016exn,Brown:2015bva,Brown:2015lvg,Fan:2018wnv,Barbon:2020olv,Barbon:2020uux,Belin:2021bga,Belin:2022xmt}, however, here we will focus on the first proposal, the so-called `complexity=volume' (CV) \cite{Susskind:2014rva,Susskind:2014jwa,Stanford:2014jda,Couch:2016exn}. Similar to the RT prescription, the CV proposal states the complexity $\mathcal{C}$ of a boundary CFT state on a Cauchy surface $\sigma_{A}$ delimiting a boundary region $A$ (such that $\partial A=\sigma_{A}$) is dual to the volume $V$ of the maximal codimension-1 bulk hypersurface $\Sigma$ homologous to $A$,\footnote{More specifically, here we consider a $d$-dimensional, compact, oriented, time-oriented Lorentzian manifold $\mathcal{M}$ with boundary $\partial \mathcal{M}$. The boundary region specifying the Cauchy slice, $A\subset\partial \mathcal{M}$, is chosen such that its causal future coincide with itself, that is, $J^+(A)=A$.}
\beq \mathcal{C}(\sigma_{A})=\frac{1}{G\ell}\,\underset{\Sigma\sim A}{\text{max}}\,V(\Sigma)\;.\label{eq:CVcomplexitydef}\eeq
Here $G$ is Newton's constant, $\ell$ is some undetermined bulk length scale, e.g., the AdS curvature scale or the horizon radius of a black hole, and the homology condition $\Sigma\sim A$ amounts to identifying the boundary of $\Sigma$ with $\sigma_{A}=\partial A$.

Due to the successes of holographic entanglement, it is natural to wonder what other aspects of gravity might be captured by holographic complexity. In fact, there is growing consensus that gravitational dynamics may arise due to complexity \cite{Czech:2017ryf,Caputa:2018kdj,Susskind:2019ddc,Pedraza:2021mkh,Pedraza:2021fgp,Pedraza:2022dqi}. Indeed, at its core, the principle of least action is, like computational complexity, fundamentally an optimization problem \cite{Pedraza:2022dqi}. A covariant realization of this sentiment was presented in \cite{Pedraza:2021mkh,Pedraza:2021fgp}, where the linearized Einstein field equations we shown to emerge from the first law of complexity, 

\beq \delta\mathcal{C}=
(\dot{\lambda}^{a}|_{\lambda_{f}})\eta_{ab}\delta\lambda^{b}_{f}\qquad \Longrightarrow \qquad \delta E_{\mu\nu}^{g}=0\;.\label{eq:firstlawcomEinintro}\eeq
More precisely, here $\mathcal{C}$ refers to a specific notion of boundary complexity that seeks the minimum number of sources $\{\lambda_{\alpha}\}$ needed to prepare a holographic CFT state from a Euclidean path integral \cite{Belin:2018fxe,Belin:2018bpg},
where $\delta\mathcal{C}$ is the variation of the complexity with respect to the sources preparing a target state $|\lambda_f\rangle$, and $\eta_{ab}$ is a metric in the auxiliary space of these sources. Combined with the CV proposal (\ref{eq:CVcomplexitydef}), which yields the holographic relation
\beq 
\delta\mathcal{C}=\frac{1}{G \ell}\delta V\;, 
\eeq
it can be shown the first law leads to the linearized Einstein equations in vacuum. Thus, gravity is a consequence of \emph{spacetime complexity}, i.e., gravitational field equations arise from spacetime minimizing the cost of computing its own dynamics \cite{Pedraza:2021mkh,Pedraza:2021fgp,Pedraza:2022dqi}. As with the other paradigms of emergent gravity, it is essential to ask whether the principle of spacetime complexity accounts for dynamics of gravitational theories beyond Einstein gravity, or even in the presence of quantum corrections.

The purpose of this article is two-fold: (i) generalize the covariant derivation of the linearized Einstein equations to arbitrary higher-derivative theories of gravity, and (ii) derive the linearized semi-classical Einstein equations. Each objective follows from an appropriate extension of the CV proposal (\ref{eq:CVcomplexitydef}) and corresponding bulk first law. In particular, we will show the linearized equations of motion of arbitrary theories of gravity follow from the boundary first law (\ref{eq:firstlawcomEinintro}) together with a proposed generalization of the CV prescription for higher derivative theories \cite{Hernandez:2020nem}
\beq
\mathcal{C}(\sigma_{A})=\frac{1}{G\ell}\,\underset{\Sigma\sim A}{\text{max}}\left[W_{\mathrm{gen}}(\Sigma)+W_{K}(\Sigma)\right]\;.
\eeq
Here $W_{\text{gen}}$ is the generalized volume functional, first considered in the context of spacetime entanglement \cite{Bueno:2016gnv}, and can be interpreted as the analog of the Iyer-Wald entropy functional for arbitrary diffeomorphism invariant theories of gravity. The second contribution, $W_{K}(\Sigma)$, consists of various contractions of the extrinsic curvature of the hypersurface $\Sigma$, and is akin to the anomaly term in the Camps-Dong formula for holographic entanglement entropy of CFTs dual to higher-derivative theories \cite{Dong:2013qoa,Camps:2013zua}. For small perturbations about vacuum AdS and for hypersurfaces with vanishing extrinsic curvature (variations about the maximal volume slice), the first law of holographic complexity becomes
\beq \delta\mathcal{C}=\frac{1}{G\ell}\delta W_{\text{gen}}\;,\eeq
and results in the linearized equations of motion for arbitrary theories of gravity. 

In the second half of this article, we study the effect of semi-classical bulk quantum corrections. We do this explicitly by evaluating the first law of holographic complexity for semi-classical Jackiw-Teitelboim (JT) gravity \cite{Jackiw:1984je,Teitelboim:1983ux}, where the problem of semi-classical backreaction can be solved \emph{exactly}. In particular, using the same set-up, we find 
\beq \delta\mathcal{C}=\frac{1}{G\ell}\delta V_{\text{JT}}+\int_{\tilde{\mathcal{M}}}\delta c_{\text{bulk}}\;,\eeq
where $\delta V_{\text{JT}}$ is the variation of the generalized volume specific to JT gravity, $\tilde{\mathcal{M}}$ is the bulk region of spacetime enclosed by $\Sigma\cup A$, and $c_{\text{bulk}}$ represents `bulk complexity' (density), i.e., the complexity due to bulk quantum fields living in AdS spacetime. The bulk correction, discussed initially in \cite{Hernandez:2020nem}, corresponds to leading $1/N$ corrections to complexity of the boundary CFT in a large-$N$ expansion, and is a formal analog to the semi-classical extension of the RT formula \cite{Faulkner:2013ana,Engelhardt:2014gca}, known as the FLM prescription.  Thus far, a precise expression for $c_{\text{bulk}}$ was lacking, with its existence only argued on general grounds. Our explicit JT gravity analysis, however, leads us to make the following proposal for the form of $\delta c_{\text{bulk}}$
\beq \delta c_{\text{bulk}}=\frac{1}{2}\delta_{Y}g_{\mu\nu}\langle \delta T^{\mu\nu}\rangle\;,\eeq
where $\delta_{Y}$ is the `new York' transformation \cite{Belin:2018bpg} of the spacetime metric $g_{\mu\nu}$, and $T^{\mu\nu}$ is the semi-classical stress-tensor of bulk matter fields. Moreover, upon invoking spacetime complexity, the quantum-corrected first law results in the linearized semi-classical Einstein equations. Hence, semi-classical gravity emerges from cost-effective computation.

The remainder of this article is outlined as follows. In Section \ref{sec:firstlawholocomp} we review holographic state preparation and its relation to complexity. We then summarize the derivation of the linearized Einstein equations via the first law of complexity. We extend this derivation to any higher-order gravity theory in Section \ref{sec:highordergravcomp} via a suitable generalization of the volume functional. In Section \ref{sec:semiclassgravcomp} we see how holographic complexity is modified by semi-classical quantum corrections. We carry out this procedure explicitly in the context of semi-classical JT gravity, where the problem of backreaction is exactly solvable. Our analysis leads us to a proposal for CV complexity including bulk quantum corrections. In Section \ref{sec:disc} we conclude with a discussion on the universality of gravity and on avenues for future work. To keep the article streamlined and self-contained we include three appendices.

\section{Holographic complexity and Einstein's equations} \label{sec:firstlawholocomp}

Here we review a specific form of the first law of complexity and how it leads to the linearized Einstein equations, as first presented in \cite{Pedraza:2021fgp,Pedraza:2021mkh}. In Section \ref{sec:highordergravcomp} we will show how to extend this derivation to arbitrary higher-order gravity theories. The first law of complexity which leads to gravitational field equations relates changes in complexity to variations of sources preparing boundary CFT states. To understand this in detail, we first review the holographic state preparation of coherent CFT states via Euclidean path integrals \cite{Skenderis:2008dh,Skenderis:2008dg,Botta-Cantcheff:2015sav,Marolf:2017kvq,Botta-Cantcheff:2019apr}.

\subsection{Holographic state preparation and a first law}\label{sec: boundaryDualOfBulkSymplecticForm}

Generally, to prepare a CFT state one performs a Euclidean path integral over the Euclidean geometry where the CFT is defined, namely, a southern hemisphere. Heuristically, preparing a coherent target state (wavefunctional)
$|\lambda_f\rangle$ from a given reference state $|\lambda_i\rangle$ is done by evaluating a (time-ordered) Euclidean path integral with sources turned on,
\beq |\lambda_f\rangle = Te^{-\int_{\tau<0}d\tau d\vec{x}\sum_{\alpha}\lambda_{\alpha}(\tau,\vec{x})\mathcal{O}_{\alpha}(\tau,\vec{x})}|\lambda_{i}\rangle\;.\label{eq:statelambdatest}\eeq 
Here $T$ refers to a time-ordering operation, $\tau$ is a Euclidean time with $\tau<0$ representing the southern hemisphere, and $\{\lambda_{\alpha}\}$ denote sources for CFT primary operators $\mathcal{O}_{\alpha}$. The reference state wavefunctional $|\lambda_i\rangle$ itself may be represented by a Euclidean path integral. In particular, when the reference state is the  CFT vacuum wavefunctional, the sources $\lambda_{i}$ are turned off and $|\lambda_i\rangle=|0\rangle\equiv\int_{\tau<0}[D\Phi]e^{-I_{E}^{\text{CFT}}}$, where $I_{E}^{\text{CFT}}$ is the CFT Euclidean action of fields $\Phi$. Similarly, conjugate states $\langle\lambda_{f}|$ are given by 
\beq 
\langle\lambda'_{f}|=\langle \lambda'_{i}| T e^{-\int_{\tau>0} d\tau d\vec{x} \lambda^{\ast}_{\alpha}\left(-\tau, \vec{x}\right) O_{\alpha}^{\dagger}\left(\tau, \vec{x}\right)},
\end{equation}
which corresponds to inserting sources $\{\lambda^{\ast}_{\alpha}\}$ (the dualization of $\lambda_{\alpha}(\tau,\vec{x})$) in the northern hemisphere. Gluing this Euclidean section to the southern hemisphere in effect computes the overlap $\langle\lambda'_{f}|\lambda_{f}\rangle$.

Already we can see how state preparation provides an intuitive description of field theory complexity, with features reminiscent of Nielsen's geometrization of circuit complexity \cite{Nielsen:2006,Nielsen:2007}. To define computational complexity, we need to associate a computational `cost' to the mapping in (\ref{eq:statelambdatest}).
The precise definition follows from recognizing that the space of coherent states $|\lambda\rangle$  is described by a manifold coordinatized by sources $\{\lambda_{\alpha}\}$.
Distances in the space of sources are given in terms of a metric $\eta_{ab}$, where the minimal path in this space is found by minimizing a cost function $F$, represented by, for example, the kinetic energy $F=\eta_{ab}\dot{\lambda}^{a}\dot{\lambda}^{b}$ \cite{Belin:2018bpg}.\footnote{There is an innate ambiguity in the definition of complexity due to a choice in cost function $F$. For example, the `geodesic distance' $F[y]=\sqrt{y}$ is a valid cost function. A motivation for choosing the kinetic energy is the expectation complexity (of tensor product states) should be additive. Moreover, a different cost function would be non-linear in $\dot{\lambda}$. This is relevant if one is interested in relating complexity to volume such that a change in volume corresponds to a linear deformation to sources \cite{Belin:2018bpg}. If, however, one does not commit themselves to CV complexity, working with other cost functions could be equally justified.}
The computational complexity $\mathcal{C}$  between a given reference state (defined by some set of sources $\lambda_{i}$) and a target state (prepared by sources $\lambda_{f}$) amounts to identifying a trajectory in the space of sources minimizing the cost function
\beq \mathcal{C}(s_{i},s_{f})=\int_{s_{i}}^{s_{f}}ds\,\eta_{ab}\dot{\lambda}_{a}\dot{\lambda}_{b}\;,\label{eq:compftdef}\eeq
with affine parameter $s$, $\lambda(s_{i,f})\equiv\lambda_{i,f}$, and $\dot{\lambda}\equiv\frac{d\lambda}{ds}$. Intuitively, the set of sources $\{\lambda_{f}\}$ act as the set of gates comprising the unitary operator transforming $|\lambda_{i}\rangle$ into $|\lambda_{f}\rangle$ in a quantum circuit. One may consider variations of the complexity with respect to $\lambda_{f}$, 
which can be used to look for variations that minimize the computational cost, i.e., 
\beq \delta_{\lambda_{f}}\mathcal{C}=(\dot{\lambda}^{a}|_{\lambda_{f}})\eta_{ab}\delta\lambda^{b}_{f}\;.\label{eq:firstlawcomv1}\eeq
Complexity thus obeys a first law. The first law (\ref{eq:firstlawcomv1}) differs from the first law of complexity presented in \cite{Bernamonti:2019zyy}, as here we consider variations with respect to the \emph{sources} that lead to a perturbative change in the target state prepared using an Euclidean path integral.\footnote{We assume a redundancy in sources such that we can prepare a target state close to the original state.} The first law of complexity in \cite{Bernamonti:2019zyy}, on the other hand, considers perturbations to the target state  corresponding to excitations of purely normalizable modes in Lorentzian signature, while non-normalizable modes (which correspond to sources of the dual operators) are kept turned off. Further, \cite{Bernamonti:2019zyy} does not make reference to any specific prescription of state preparation.

Let us explain (\ref{eq:firstlawcomv1}) more precisely. The space of coherent states is described by a K{\"a}hler (and hence symplectic) manifold with K{\"a}hler potential $\mathcal{K}$ and symplectic 2-form $\Omega_{\text{bdry}}$,
\beq \mathcal{K}=\log\langle\lambda|\lambda\rangle\;,\quad  \Omega_{\text{bdry}}=i\partial_{\lambda_{\alpha}}\partial_{\lambda^{\ast}_{\alpha'}}\log\langle\lambda|\lambda\rangle d\lambda_{\alpha}\wedge d\lambda_{\alpha'}^{\ast}\;,\eeq
and arbitrary coherent state $|\lambda\rangle$.\footnote{As a coherent state, $|\lambda\rangle$ is unnormalized. The set of coherent states is complete but not orthogonal.}  Denoting global coordinates on the space of sources by $\tilde{\lambda}=(\lambda,\lambda^{\ast})$, the symplectic form may be cast in terms of variations of $\tilde{\lambda}$ \cite{Belin:2018fxe}
\beq \Omega_{\text{bdry}}(\delta_{1}\tilde{\lambda},\delta_{2}\tilde{\lambda})=i\partial_{\lambda_{\alpha}}\partial_{\lambda_{\alpha'}^{\ast}}\log\langle\lambda|\lambda\rangle[\delta_{1}\lambda_{\alpha}\delta_{2}\lambda_{\alpha'}^{\ast}-\delta_{2}\lambda_{\alpha}\delta_{1}\lambda_{\alpha'}^{\ast}]\;.\label{eq:Ombdryv1}\eeq
In the case of coherent CFT states, the K{\"a}hler potential is related to the CFT partition function $Z_{\text{CFT}}[\tilde{\lambda}]\equiv\langle \lambda|\lambda\rangle$ via $\mathcal{K}=\log Z_{CFT}[\tilde{\lambda}]$, where $\mathcal{K}$ is understood as a functional of half-sided sources $(\lambda,\lambda^{\ast})$, and the symplectic form  (\ref{eq:Ombdryv1}) becomes
\beq \Omega_{\text{bdry}}(\delta_{1}\tilde{\lambda},\delta_{2}\tilde{\lambda})=i(\delta_{1}^{\ast}\delta_{2}-\delta_{2}^{\ast}\delta_{1})\log Z_{CFT}[\tilde{\lambda}]\;. \label{eq: boundarysymplecticform}\eeq
Returning to the first law (\ref{eq:firstlawcomv1}), there exist special deformation of the sources, $\delta_{Y}\lambda$, such that $\eta(\delta_{Y}\lambda,\delta\lambda)=\Omega_{\text{bdry}}(\delta_{Y}\lambda,\lambda)$ \cite{Belin:2018fxe}; more precisely, the special deformations satisfy $\dot{\lambda}^{a}|_{\lambda_{f}}=J[\delta_{Y}\lambda]$, where $J$ is the complex structure relating the metric $\eta$ and symplectic form $\Omega_{\text{bdry}}$, where we recall that in general a K{\"a}hler metric is related to its K{\"a}hler form via $\eta(\delta\lambda,\delta\lambda)=\Omega(\delta\lambda,J[\delta\lambda])$. Thence, for such deformations the first law (\ref{eq:firstlawcomv1}) is
\beq \delta_{\lambda_{f}}\mathcal{C}=\Omega_{\text{bdry}}(\delta_{Y}\lambda,\delta\lambda)\;.\label{eq:firstlawcomnograv}\eeq
It is worth stressing this is purely a field theory statement.

\begin{figure}[t]
$\qquad\qquad\quad$\includegraphics[scale=0.45,trim=0 2cm 10cm 12cm]{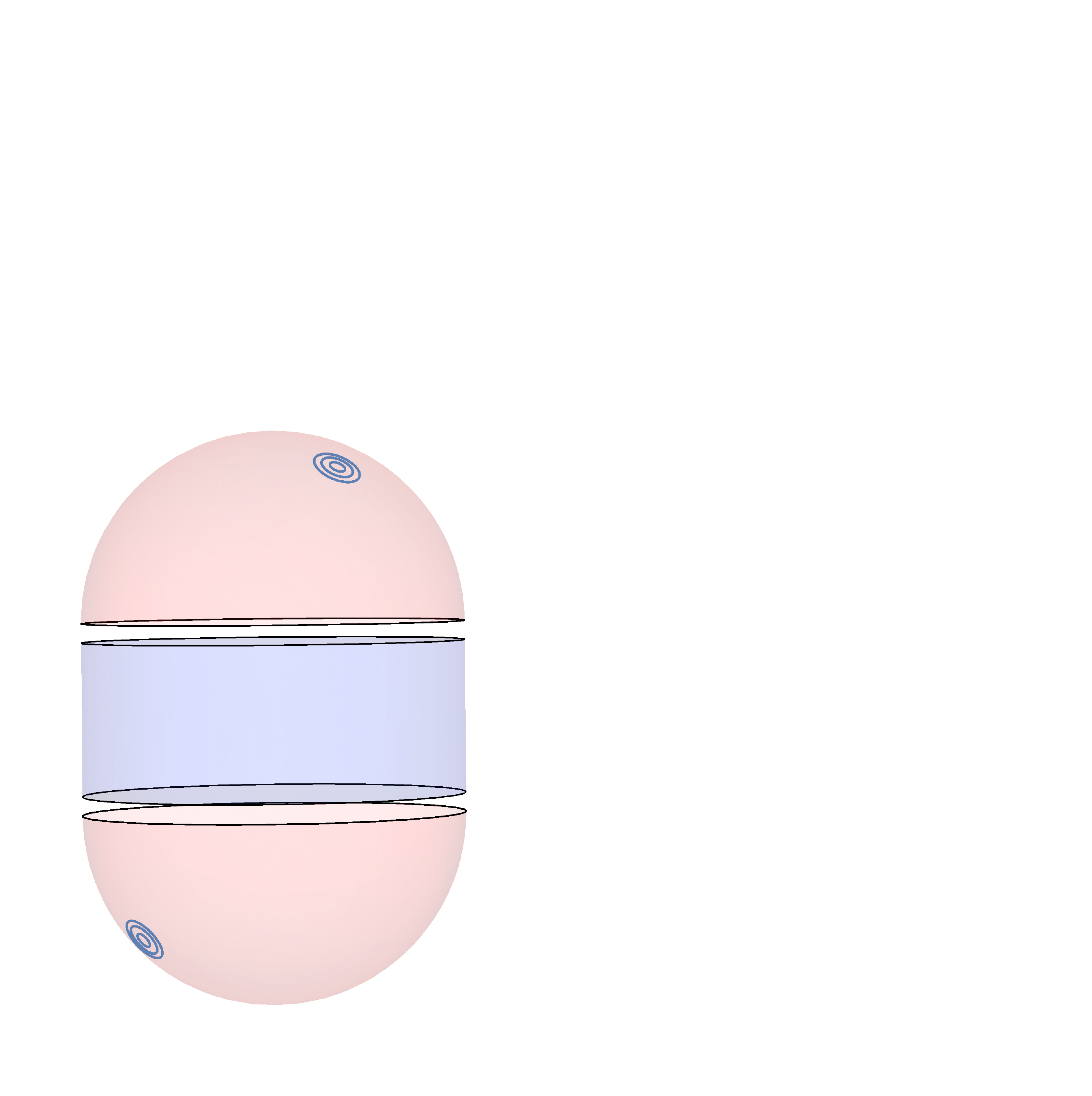}
\centering
\begin{picture}(0,0)
\put(-222,19){$\lambda_f$}
\put(-175,23){$\mathcal{M}_{-}$}
\put(-143,203){$\lambda_f'$}
\put(-120,62){$\Sigma_{-}$}
\put(-120,86){$\Sigma_{-}$}
\put(-175,100){$\tilde{\mathcal{M}}$}
\put(-225,123){$\Sigma_{+}$}
\put(-225,147){$\Sigma_{+}$}
\put(-178,171){$\mathcal{M}_{+}$}
\put(-96,75){$|\lambda_f\rangle$}
\put(-96,138){$\langle\lambda'_f|$}
\end{picture}
\caption{Holographic state preparation of coherent CFT states. The sources and the reference state defined on the southern hemisphere of $\mathcal{M}_{-}$ prepare the target state on $\Sigma_{-}$. Given initial analytic data on $\Sigma_{-}$, Einstein's equations describe Lorentzian evolution in $\tilde{\mathcal{M}}$, ending at the slice $\Sigma_{+}$. A complete transition amplitude requires gluing the Euclidean submanifold $\mathcal{M}_{+}$ to $\Sigma_{+}$, closing the contour of integration.
}
\label{fig: prepforlorentzian2}
\end{figure}
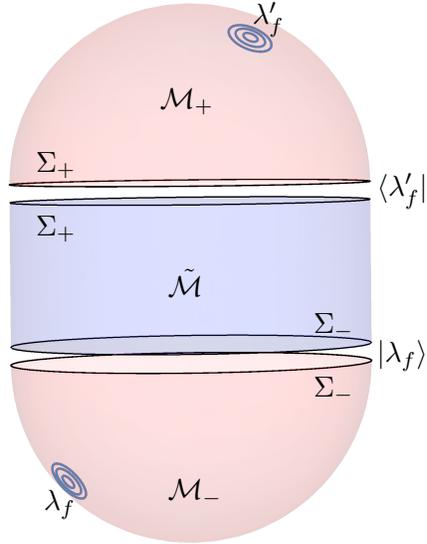

In the context of the $\text{AdS}_{d+1}/\text{CFT}_{d}$ correspondence, preparation of coherent CFT states is described by a path integral over the boundary Euclidean AdS $\mathcal{M}$, where the southern hemisphere has topology $S^{d-1}\times\mathbb{R}^{-}\sim B^{d}$, which we denote by $\partial\mathcal{M}_{-}$. Holographically speaking, CFT state preparation maps to the preparation of a semi-classical bulk gravitational state on a bulk Cauchy slice $\Sigma_{-}$ \cite{Skenderis:2008dh,Skenderis:2008dg,Botta-Cantcheff:2015sav,Marolf:2017kvq,Botta-Cantcheff:2019apr}. More specifically, via the standard AdS/CFT dictionary, the boundary values of the bulk fields in a southern Euclidean AdS submanifold $\mathcal{M}_{-}$ specify the reference state $|\lambda_i\rangle$ and the sources $\lambda_f$ used to prepare the target state $|\lambda_{f}\rangle$. When the bulk fields are on-shell, the boundary values of the fields uniquely determine their values on the Cauchy slice $\Sigma_{-}$, representing the target state $|\lambda_f\rangle$. The time evolution of the CFT state then follows from solving the bulk Einstein's equations with such initial data, which is represented by a section of a Lorentzian cylinder, as depicted in Figure \ref{fig: prepforlorentzian2}. The contour may be closed by gluing the northern Euclidean AdS submanifold $\mathcal{M}_{+}$ to the Lorentzian cylinder.

A key observation made in \cite{Belin:2018fxe} is that the mapping between boundary sources and initial data persists at the level of the respective symplectic structures. Namely, the symplectic form $\Omega_{\text{bulk}}(\phi,\delta_{1}\phi,\delta_{2}\phi)$ on the classical phase space of bulk dynamical field configurations $\phi$ is dual to the symplectic form $\Omega_{\text{bdry}}(\delta_{1}\tilde{\lambda},\delta_{2}\tilde{\lambda})$ characterizing the space of sources. Specifically, when the boundary CFT is holographic, the standard AdS/CFT dictionary states
\beq Z_{\text{CFT}}[\tilde{\lambda}]=\langle\lambda|\lambda\rangle=e^{-I_{E,\text{grav}}^{\text{on-shell}}[\tilde{\lambda}]}\;,\eeq
with $\mathcal{K}=-I_{E,\text{grav}}^{\text{on-shell}}$ being the on-shell bulk gravity action, and where $\tilde{\lambda}$ set the boundary conditions for the bulk fields $\phi$ according to the prescription of piece-wise holography \cite{Skenderis:2008dh,Skenderis:2008dg}. Consequently, via the boundary symplectic form (\ref{eq: boundarysymplecticform}), one finds \cite{Belin:2018fxe} 
\beq \Omega_{\text{bdry}}(\delta_{1}\tilde{\lambda},\delta_{2}\tilde{\lambda})=i(\delta_{2}^{\ast}\delta_{1}-\delta_{1}^{\ast}\delta_{2})I_{\text{grav}}^{\text{on-shell}}[\lambda,\lambda^{\ast}]=i\int_{\partial\mathcal{M}_{-}}\omega_{\text{bulk}}^{E}(\phi,\delta_{1}\phi,\delta_{2}\phi)\;,\label{eq:bulkbdrysymps}\eeq
where $\omega_{\text{bulk}}^{E}$ refers to the bulk symplectic current form in Euclidean signature evaluated at the boundary of Euclidean AdS, and the extrapolate dictionary is used to relate the sources $\tilde{\lambda}$ to the boundary values of the dual bulk fields $\phi$. 

It is worth pausing here for a moment to briefly review how this equivalence between bulk and boundary symplectic forms is established \cite{Belin:2018fxe}. The result follows from an application of the covariant phase space formalism as developed by \cite{Lee:1990nz,Wald:1993nt,Iyer:1994ys,Wald:1999wa}. Let $\mathcal{M}$ be a $d$-dimensional Euclidean spacetime endowed with a Euclidean metric $g$ and consider a Lagrangian field theory that is covariant under arbitrary diffeomorphisms. The Lagrangian $d$-form $L$ is solely a function of dynamical fields $\phi$ (and its derivatives). Under infinitesimal field variations, the variation of the Lagrangian is
\beq \delta L=-E_{\phi}\delta\phi+d\theta(\phi,\delta\phi)\;,\label{eq:Lformvar}\eeq
where $E_{\phi}$ is the equations of motion $d$-form, with an implicit sum over dynamical fields, and $\theta(\phi,\delta\phi)$ is the  symplectic\footnote{Technically, at this stage $\theta$ is the pre-symplectic potential as the space of kinematically allowed field configurations does not constitute a physical phase space, a symplectic manifold with a non-degenerate symplectic form. For our purposes, this distinction is unimportant.} potential $(d-1)$-form, which is locally build from $\phi$, $\delta\phi$ and their derivatives and is linear in field variations $\delta\phi$. For on-shell field configurations, $E_{\phi}=0$, it follows the variation of the on-shell Euclidean gravity action is a boundary term, 
\beq \delta I_{E,\text{grav}}^{\text{on-shell}}[\tilde{\lambda}]=\int_{\mathcal{M}}\delta L=\int_{\partial\mathcal{M}=S^{d}}\theta(\tilde{\lambda},\delta\tilde{\lambda})\;,\eeq
where the extrapolate dictionary was used. With the joint source profile $\tilde{\lambda}=(\lambda,\lambda^{\ast})$, and restricting ourselves to holomorphic or anti-holomorphic variations, the integral over $\partial\mathcal{M}$ will localize on the northern and southern hemispheres, respectively, such that 
\beq 
\begin{split}
 \Omega_{\text{bdry}}(\delta_{1}\tilde{\lambda},\delta_{2}\tilde{\lambda})&=
i\int_{\partial\mathcal{M}_{-}}\omega_{\text{bulk}}^{E}(\phi,\delta_{1}\phi,\delta_{2}\phi),
\end{split}
\eeq
where $\omega_{\text{bulk}}^{E}$ is the (Euclidean) bulk symplectic current $(d-1)$-form, defined as
\beq \omega(\phi,\delta_{1}\phi,\delta_{2}\phi)\equiv\delta_{1}\theta(\phi,\delta_{2}\phi)-\delta_{2}\theta(\phi,\delta_{1}\phi)\;.\eeq
 The symplectic current is locally constructed out of fields $\phi$, its field variations, and their derivatives, and is linear in $\delta_{1,2}\phi$ and its derivatives.

In deriving the equivalence (\ref{eq:bulkbdrysymps}), we emphasize the bulk dynamical fields are on-shell, $E_{\phi}=0$. Notably, moreover, the integral in $\Omega_{\text{bdry}}$ is over the southern hemisphere, however, recall that the bulk symplectic form $\Omega_{\text{bulk}}(\phi,\delta_{1}\phi,\delta_{2}\phi)$ is defined as the integral over a Cauchy slice $\Sigma$ \cite{Lee:1990nz,Wald:1993nt,Iyer:1994ys,Wald:1999wa}. This is a consequence of the fact  $\omega_{\text{bulk}}$ defines a conserved current when the field variations $\delta_{1,2}\phi$  obey the linearized equations of motion, $\delta_{1,2}E_{\phi}=0$, resulting in $d\omega_{\text{bulk}}=0$.\footnote{Explicitly, $d\omega=\delta_{1}d\theta(\phi,\delta_{2}\phi-\delta_{2}d\theta(\phi,\delta_{1}\phi)=(\delta_{1}E_{\phi})\delta_{2}\phi-(\delta_{2}E_{\phi}\delta_{1}\phi$, where we used that the spacetime exterior derivative $d$ and field variation commute, and field variations commute such that $\delta_{1}\delta_{2}L=\delta_{2}\delta_{1}L$. Hence $\omega$ is a closed $(d-1)$-form on spacetime, which, defines a covariantly conserved current $\omega^{\mu}$, $\nabla_{\mu}\omega^{\mu}=0$.} Consequently, $\omega_{\text{bulk}}^{E}$ can be `pushed' to other codimension-1 hypersurfaces. In particular, the southern hemisphere $\partial\mathcal{M}_{-}$ is pushed to a $\tau=0$ surface $\Sigma$ in $\mathcal{M}$. The slice $\Sigma$ analytically continues to a Lorentzian initial value surface (at $t=0$ and also denoted as $\Sigma$) when bulk configurations, i.e., $\tilde{\lambda}$ are $\mathbb{Z}_{2}$ symmetric, and one restricts to variations in the complexified tangent space which correspond to real Lorentzian initial data.\footnote{More precisely, the configuration is always symmetric under a time-reversal \emph{plus} complex conjugation. The fixed-surface of that symmetry, however, may not be at $t=\tau=0$. This only
occurs if the sources are purely real, where all the momenta vanish on the continuation slice.} In so doing, the holographic dual of the boundary symplectic form is dual to the bulk symplectic form \cite{Belin:2018fxe}
\beq \Omega_{\text{bdry}}(\delta_{1}\tilde{\lambda},\delta_{2}\tilde{\lambda})=\int_{\Sigma}\omega_{\text{bulk}}(\phi,\delta\phi_{1},\delta\phi_{2})=\Omega_{\text{bulk}}(\phi,\delta_{1}\phi,\delta_{2}\phi)\;.
\label{eq: boundaryEqualsInitial}\eeq
Here $\omega_{\text{bulk}}$ is the Lorentzian  symplectic current form.

\subsubsection*{The `new York' time transformation}

As mentioned above, there exist special deformations of the sources $\delta_{Y}\lambda$ such that the first law of complexity takes the form (\ref{eq:firstlawcomnograv}).  Holographically, the first law becomes, via (\ref{eq: boundaryEqualsInitial}),
\beq \delta_{\lambda_{f}}\mathcal{C}=\Omega_{\text{bdry}}(\delta_{Y}\tilde{\lambda},\delta\tilde{\lambda})=\Omega_{\text{bulk}}(\phi,\delta_{Y}\phi,\delta\phi)\;,\eeq
when variations $\delta_{Y}\phi$ and $\delta\phi$ obey the linearized equations of motion, $\delta_{Y}E_{\phi}=\delta E_{\phi}=0$. Assuming the CV conjecture, it was shown in \cite{Belin:2018fxe,Belin:2018bpg} these special deformations $\delta_{Y}$ are on-shell when it occurs on the maximal volume slice. Then, via the bulk symplectic form, variations $\delta_{Y}$ correspond to variations in volume $V$ of extremal bulk hypersurfaces, such that 
\beq \delta_{\lambda_{f}}\mathcal{C}=\Omega_{\text{bdry}}(\delta_{Y}\tilde{\lambda},\delta\tilde{\lambda})=\Omega_{\text{bulk}}(\phi,\delta_{Y}\phi,\delta\phi)=\frac{1}{G\ell}\delta V\;.\label{eq:holofirstlawv1}\eeq
 Note that the holographic first law (\ref{eq:holofirstlawv1}) is consistent with CV duality, though, it does not constitute a proof of the CV conjecture as the first law only refers to variations of the complexity. Moreover, here we have a specific notion of boundary complexity in mind, Eq. (\ref{eq:compftdef}), such that one attains the change in volume on the extremal slice $\Sigma$. In principle, however, one could consider a different cost function whose variation would be dual to a variation of some other  bulk geometric functional. We will return to this point in Section \ref{sec:disc}. 
 

 The special deformation $\delta_{Y}$ is dubbed the ``new York'' transformation \cite{Belin:2018bpg} due to its similarities with York time \cite{York:1972sj}. Since we will see how this transformation leads to a generalized volume functional in the case of higher-order gravities, let us briefly summarize its character in the context of (Lorentzian) Einstein gravity. 
 
 It is convenient to work with the Arnowitt-Deser-Misner (ADM) formalism (see Appendix \ref{app:ADMforms} for a relevant review). Let $\tilde{\mathcal{M}}$ denote a $d$-dimensional Lorentzian bulk spacetime with local coordinates $x^{\mu}$ ($\mu=0,...,d-1$). Foliate the bulk by codimension-1 hypersurfaces $\Sigma_{t}$ of constant time $t$, with a timelike unit normal $n^{\mu}$ and coordinatized by $y^{a}$ ($a=1,...,d-1$). The  induced metric on $\Sigma_{t}$ is $h_{\mu\nu}= g_{\mu\nu} + n_{\mu} n_{\nu}$ and the extrinsic curvature of $\Sigma_{t}$ is $K_{\mu\nu}=h^{\lambda}_{\mu}\nabla_{\lambda}n_{\nu}$. Projected onto $\Sigma_{t}$, the induced metric and extrinsic curvature are, respectively, $h_{ab}=e^{\mu}_{a}e^{\nu}_{b}h_{\mu\nu}$ and $K_{ab}=e^{\mu}_{a}e^{\nu}_{b}K_{\mu\nu}$, with $e^{\mu}_{a}\equiv\frac{\partial x^{\mu}}{\partial y^{a}}$. The ADM action $I_{\text{ADM}}$ for vacuum Einstein gravity is (with $16\pi G=1$)
\beq I_{\text{ADM}}=\int dt \int_{\Sigma_{t}}d^{d-1}y\mathcal{L}_{\text{ADM}}+I_{\mathcal{B}}\;,\eeq
supplemented by an appropriate boundary term on the asymptotic timelike boundary $\mathcal{B}$. Here $\mathcal{L}_{\text{ADM}}$ is the ADM Lagrangian density $\mathcal{L}_{\text{ADM}}=\pi^{ab}\dot{h}_{ab}-2\Lambda-\mathcal{H}_{\text{ADM}}$, with conjugate momenta
\beq \pi^{ab}=\sqrt{h}(K^{ab}-Kh^{ab})\;,\eeq
and $\mathcal{H}_{\text{ADM}}=\sqrt{h}(N\mathcal{H}+N_{a}\mathcal{H})$ is the ADM Hamiltonian density, where $N$ and $N_{a}$ denote the lapse and shift, respectively, while $\mathcal{H}$ and $\mathcal{H}_{a}$ are the Hamiltonian and momentum constraints,
\beq 
\begin{split}
0=\mathcal{H}&=-(\bar{R}-2\Lambda)+K_{ab}^{2}-K^{2}=-(\bar{R}-2\Lambda)+\frac{1}{h}\left(\pi_{ab}\pi^{ab}-\frac{\pi^{2}}{(d-2)}\right)\;,
\end{split}
\label{eq:HamconstGR}\eeq
\beq 0=\mathcal{H}^{a}=-2\nabla_{b}\left(h^{-1/2}\pi^{ab}\right)\;.\label{eq:momconstGR}\eeq
Here $\bar{R}$ is the Ricci scalar of $h_{ab}$ and $\pi\equiv h^{ab}\pi_{ab}$. The momentum constraint is associated with diffeomorphisms inside $\Sigma_{t}$, while the Hamiltonian constraint is related to diffeomorphisms which change the initial value surface.

Varying the ADM action yields
\beq \delta I_{\text{ADM}}=\int_{\Sigma_{t}}d^{d-1}y\,\pi^{ab}\delta h_{ab}\;\;+\;\;\text{bulk EOM}\;.\eeq
Comparing to the variation of Lagrangian form (\ref{eq:Lformvar}) (in Lorentzian signature) such that $\delta I_{\text{ADM}}=\int_{\tilde{\mathcal{M}}}E_{\phi}\delta\phi+\int_{\Sigma_{t}}\theta(\phi,\delta\phi)$, we read off the symplectic potential form to be $\theta(g,\delta g)=\pi^{ab}\delta h_{ab}$. Hence, the corresponding bulk symplectic form is \cite{Lee:1990nz}
\beq \Omega_{\text{bulk}}(g,\delta_{1}g,\delta_{2}g)=\int_{\Sigma_{t}}(\delta_{1}\pi^{ab}\delta_{2}h_{ab}-\delta_{2}\pi^{ab}\delta_{1}h_{ab})\;.\label{eq:bulksympformGR}\eeq
To be on-shell, one must be on the constraint submanifold where the pair $(h_{ab},\pi^{ab})$ satisfy the constraints (\ref{eq:HamconstGR}) and (\ref{eq:momconstGR}).

Arbitrary $(h_{ab},\pi^{ab})$  may not be `good' phase space\footnote{In the sense that by phase space one means the submanifold where the constraints of the theory are satisfied, i.e., the set of solutions to the classical equations of motion.} variables since the Hamiltonian constraint is not always solvable (alternately, the momentum constraint may always be satisfied by fixing a gauge). However, York showed there is a general procedure for solving the Hamiltonian constraint (\ref{eq:HamconstGR}) when one provides initial data on a surface with constant mean curvature \cite{York:1972sj}. The essential idea is to  separate the induced metric into a scale captured by the volume element $\sqrt{h}$ on $\Sigma_{t}$, and a conformal metric $\bar{h}_{ab}=|h|^{-1/(d-1)}h_{ab}$. In these new variables, one has $\pi^{ab}\delta h_{ab}=\pi_{V}\delta\sqrt{h}+\bar{\pi}^{ab}\delta \bar{h}_{ab}$, with\footnote{Explicitly, it is straightforward to show $\pi^{ab}\delta h_{ab}=\pi_{V}\delta\sqrt{h}+h^{\frac{1}{(d-1)}+\frac{1}{2}}(K_{ab}-Kh^{ab})\delta\bar{h}_{ab}$. However, note that $h^{ab}\delta\bar{h}_{ab}$, a consequence of the fact that, by definition $\delta \sqrt{\bar{h}}=0$. Hence, we may add in any constant function in the second term. The choice made is such because, since the conjugate momenta $\pi_{V}$ to the volume density $\sqrt{h}$ is proportional to the trace of the extrinsic curvature, then the conjugate momenta to the conformal metric $\bar{h}_{ab}$ must be proportional to the traceless part of the extrinsic curvature.}
\beq \pi_{V}=-\frac{2(d-2)}{(d-1)}K\;,\quad \bar{\pi}^{ab}=|h|^{\frac{1}{(d-1)}+\frac{1}{2}}\left(K^{ab}-\frac{1}{(d-1)}Kh^{ab}\right)\;.\label{eq:Yorkvarbs}\eeq
When $\pi_{V}$ is constant, the Hamiltonian constraint (\ref{eq:HamconstGR}) may be interpreted as a differential equation in the volume density $\sqrt{h}$, known as the Lichnerowitz equation. This equation is in fact solvable in flat and AdS space, admitting a unique solution such that the volume density may be cast as a functional of the remaining phase space variables,  $\sqrt{h[\bar{h},\pi_{V},\bar{\pi}]}$. Working in a constant mean curvature (CMC) slicing, where each slice has constant $K$, $\pi_{V}$ is a number parametrizing each of the slices and can be interpreted as time -- the York time. Meanwhile, the volume $V=\int\sqrt{h}$ can be understood as a Hamiltonian.\footnote{This can be understood by comparing to classical mechanics, where one considers on-shell variations in the particle trajectory $q$, variations of time $t$, and makes use of the Hamilton-Jacobi equation, such that the symplectic form lives in an `extended phase space'. See \cite{Belin:2018bpg} for more details.}

The decomposition by York can be used to provide a boundary interpretation of the volume $V$ \cite{Belin:2018fxe,Belin:2018bpg}. Specifically, the `new York' deformation $\delta_{Y}$ of the Euclidean boundary data gives rise to a change in volume of maximal slices $\Sigma$ via the duality of boundary and bulk symplectic forms (\ref{eq: boundaryEqualsInitial}). Explicitly, one fixes the following variations
\beq \delta_{Y}\pi_{V}=2(d-2)\alpha\;,\quad \delta_{Y}\bar{\pi}^{ab}=\delta_{Y}\sqrt{h}=\delta_{Y}\bar{h}_{ab}=0\;,\label{eq:Yorkvariations}\eeq
where $\alpha$ is some constant. Consequently, from the new variables (\ref{eq:Yorkvarbs}) one finds
\begin{equation} \label{eq: yorkTransformation}
\delta_{Y}h_{ab}=0\;,\quad \delta_{Y} K_{ab}=-\alpha h_{ab}\;.
\end{equation}
Moreover, in terms of variables (\ref{eq:Yorkvarbs}), the bulk symplectic form (\ref{eq:bulksympformGR}) becomes
\beq \Omega_{\text{bulk}}(g,\delta_{Y}g,\delta g)=\int_{\Sigma_{t}}\delta_{Y}(\pi^{ab}\delta h_{ab})=\int_{\Sigma_{t}}(\delta_{Y}\pi_{V})\delta\sqrt{h}=\frac{(d-2)\alpha}{8\pi G}\delta V\;,\label{eq:OmbulkdeltaV}\eeq
where we implemented the variations (\ref{eq:Yorkvariations}) and used that variations commute, and in the last line we restored $G$.

In general, however, the new York deformation (\ref{eq: yorkTransformation}) is not on-shell: the Hamiltonian constraint is generally not preserved under $\delta_Y$.\footnote{The momentum constraint is automatically satisfied: $\delta_{Y}\mathcal{H}^{a}=-2\nabla_{b}(\delta_{Y}K^{ab}-h^{ab}\delta K)=-2(d-2)\alpha\nabla_{b}h^{ab}=0$, where $\delta_{Y}\nabla_{b}$ is zero since it only depends on tangential derivatives of the metric and $\nabla_{b}h^{ab}=0$.} Specifically, the variation of (\ref{eq:HamconstGR}) is
\begin{equation}
\delta_Y \mathcal{H} = 2 \alpha( d-2) K\;.
\end{equation}
where we implemented (\ref{eq: yorkTransformation}). Thus, $\delta_{Y}$ is an on-shell perturbation provided the deformation occurs on a maximal slice $\Sigma$, where $K=0$. In other words, $\delta_{Y}$ is on-shell when $V$ is the volume of the maximal hypersurface $\Sigma$. In choosing $\alpha$ such that the coefficient becomes $(8G\ell)^{-1}$, we see by CV duality that $\Omega_{\text{bulk}}(\delta_{Y}g,\delta g)$ encodes a notion of varying complexity, $\delta\mathcal{C}$. More carefully, by the equivalence of boundary and bulk symplectic forms, $\delta V$ is equal to the boundary symplectic form (on $\partial\mathcal{M}$), resulting in the holographic first law (\ref{eq:holofirstlawv1}).

Note that the new York transformation (\ref{eq: yorkTransformation}) is generically not a diffeomorphism since it does not evolve the gauge invariant initial data $(\bar{\pi},\bar{h}$) in (York) time. Rather $\delta_{Y}$ copies the initial data to a neighboring slice. However, for deformations about empty AdS, the new York transformation is in fact a diffeomorphism \cite{Belin:2018bpg} (see Appendix \ref{app:NYVaccumAdS} for a review).

\subsection{Einstein's equations from spacetime complexity} 

\begin{figure}[t!]
\centering
\begin{tikzpicture}[scale=.9]
	\pgfmathsetmacro\myunit{4}
	\draw[white]	(0,0)			coordinate (a)
		--++(90:\myunit)	coordinate (b);
	\draw [white] (b) --++(0:\myunit)		coordinate (c)
							node[pos=1, above] {\color{black} $\Omega_{\text{bdry}}$};
                            
    \draw[white] (c) --++(-90:\myunit)	coordinate (d);                        
    \draw[line width=0.45mm] (a) -- (d) ;
    \draw[line width=0.45mm] (d) --++(0:\myunit)	coordinate (e);
    \draw[line width=0.45mm] (a) -- (e) node[pos=.5, below] {$\text{complexity=volume}$};
    \draw[line width=0.45mm] (c) -- (e) node[pos=1, below] {$\delta V$}
                                node[pos=.5, above, sloped] {$\text{Linearized EOM}$};
    \draw[line width=0.45mm] (a) -- (c) node[pos=0, below] {$\delta \mathcal{C}$}
                        node[pos=.5, above, sloped] {$\text{Boundary first law}$};
\end{tikzpicture}
\caption{A trinity between holographic complexity $\mathcal{C}\sim V$, a boundary first law $\delta\mathcal{C}=\Omega_{\text{bdry}}$, for complexity given in (\ref{eq:compftdef}), and the equivalence between boundary symplectic form and change in volume of the maximal hypersurface via the bulk symplectic form evaluated over on-shell deformations. 
}
\label{fig:trinity}  
\end{figure}
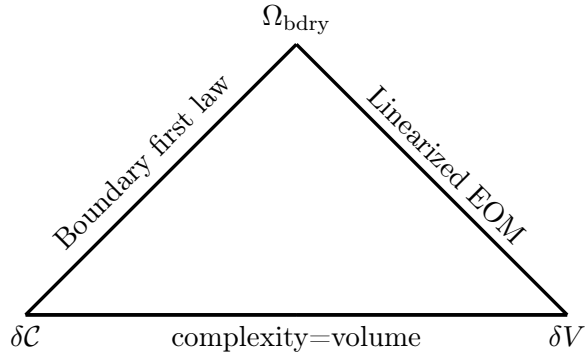

Above we reviewed the equivalence between boundary and bulk symplectic forms (\ref{eq: boundaryEqualsInitial}), for arbitrary on-shell variations. On the boundary, the new York deformation of the sources shows varying complexity (where complexity takes the specific form in (\ref{eq:compftdef})) is equivalent to the boundary symplectic form (\ref{eq:firstlawcomnograv}).\footnote{This amounts to identifying $J[\delta_{Y}\lambda]^{a}=\dot{\lambda}^{a}|_{\lambda_{f}}$, where $J[\delta_{Y}\lambda]$ is the complex structure compatible with the symplectic structure of the space of sources, i.e., $\Omega(\delta_{Y}\lambda,\delta \lambda)=\eta_{ab}J[\delta_{Y}\lambda]^{a}\delta\lambda^{b}$.} In the bulk, the symplectic form associated with on-shell new York perturbations is proportional to the variation of the volume of the extremal slice (\ref{eq:OmbulkdeltaV}). Combining each element, and assuming the complexity=volume proposal, results in the holographic first law of complexity (see Figure \ref{fig:trinity} for a visualization).

In \cite{Pedraza:2021mkh,Pedraza:2021fgp} it was shown the linearized Einstein equations of motion arise from imposing two legs of the triangle depicted in Figure \ref{fig:trinity}. Namely, assuming the boundary first law and complexity=volume, the holographic first law implies (setting $\alpha=8\pi/(d-2)\ell$) 
\beq \frac{1}{G \ell}\delta V=\Omega_{\text{bdry}}(\delta_{Y}\tilde{\lambda},\delta\tilde{\lambda})\Rightarrow \delta E_{\mu\nu}=0\;,\label{eq:firstlawassumption}\eeq
where $\delta E_{\mu\nu}=0$ denotes the linearized Einstein's equations for perturbations about vacuum AdS. The starting point of the derivation is the following application of Stokes' theorem
\begin{equation}\label{eq: integralofdomega0}
i \int_{\mathcal{M}_{-}} d \omega^{E}_{\text{bulk}}(g,\delta_{Y}g,\delta g) = i\left( \int_{\partial \mathcal{M}_{-}} \omega^{E}_{\text{bulk}}(g,\delta_{Y}g,\delta g) - \int_{\Sigma} \omega^{E}_{\text{bulk}}(g,\delta_{Y}g,\delta g)\right)\;,
\end{equation}
which must  hold for all variations that yield real Lorentzian initial data on $\Sigma$. It is sufficient to consider variations $\delta = \delta^+ + \delta^-$, where $\delta^{\pm}$ localize to $\mathcal{M}_{\pm}$, and agree on $\Sigma$.\footnote{For variations  $\delta^{\pm}$ to agree on $\Sigma$ requires $\delta^{+} g|_{ \Sigma}  =\delta^{-} g|_{ \Sigma}$ and $\partial_t (\delta \phi^{+} )^{\ast}|_{\Sigma}  = - \partial_t ( \delta \phi^{-})|_{ \Sigma}$. Consequently, `good' Lorentzian initial data $(\varphi_{L},\pi_{L})$ can be defined by $\delta \varphi_L = \text{Re} [ (\delta^+ \phi)|_{ \Sigma} ]$ and $\delta \pi_L = \text{Im} [ (\partial_t \delta^+ \phi)|_ \Sigma]$.} 
The contribution over $\partial \mathcal{M}_{-}$ is identified as the boundary symplectic form $\Omega_{\text{bdry}}(\delta_{Y}\tilde{\lambda},\delta\lambda)$, while the $\Sigma$ integral is $\Omega_{\text{bulk}}(g,\delta_{Y}g,\delta g)=(\ell G)^{-1}\delta V$, where we perturb around Lorentzian initial data. Thus,
\begin{equation}\label{eq: pushingVolumeToBoundary}
i\int_{ \mathcal{M}_{-}} d \omega^{E}_{\text{bulk}}( g,\delta_Y g, \delta g) =\Omega_{\text{bdry}}( \delta_Y \tilde{\lambda}, \delta\tilde{\lambda}) - \frac{1}{G \ell}\delta V\;.
\end{equation}
Note that at this stage the right-hand side does not vanish automatically. When it does, namely, when the first law of holographic complexity holds, it follows $d \omega_{\text{bulk}}^{E}( \delta_Y g, \delta g)$ must vanish in all of $\mathcal{M}_{-}$. A similar analysis can be done for the northern hemisphere such that $d \omega_{\text{bulk}}^{E}( g,\delta_Y g, \delta g)=0$ in all of $\mathcal{M}_+$. Thus, $d \omega_{\text{bulk}}^{E}( g,\delta_Y g, \delta g)$ must vanish everywhere in $\mathcal{M}$. 

Since the symplectic current form $\omega_{\text{bulk}}$ is closed over spacetime when the field variations obey the linearized equations of motion (see above Eq. (\ref{eq: boundaryEqualsInitial})), it is not surprising that enforcing the right-hand side of (\ref{eq: pushingVolumeToBoundary}) to vanish implies the linearized Einstein equations of motion. To see this, consider $d\omega_{\text{bulk}}^{E}(g,\delta_{Y}g,\delta g)$ in vacuum AdS
\beq d\omega_{\text{bulk}}^{E}(g,\delta_{Y}g,\delta g)= \delta_Y E^{\mu\nu} \delta g_{\mu\nu} - \delta E^{\mu\nu} \delta_Y g_{\mu\nu}\;,
\label{eq: bulkintegral}
\end{equation}
where we invoked the variation of the Lagrangian (\ref{eq:Lformvar}) and that variations commute. Since the new York deformation $\delta_{Y}$ is a diffeomorphism for perturbations around vacuum AdS,  it follows $\delta_Y E_{\mu\nu} =0$ and \eqref{eq: bulkintegral} reduces to\footnote{More generally, $\delta_{Y}$ is not a diffeomorphism on arbitrary backgrounds, however, $\delta_{Y}$ is on-shell, such that $\delta_{Y}E_{\mu\nu}=0$ for perturbations around any on-shell background.}
\begin{equation}
d \omega_{\text{bulk}}^{E}(g,\delta_Y g, \delta g)=- \delta E^{\mu \nu} \delta_Y g_{\mu \nu}\;.
\label{eq:domEIn}\end{equation}
Hence,  demanding $\int_{\mathcal{M}} d\omega^{E}_{\text{bulk}}(g,\delta_Y g, \delta g)$ vanishes for all variations amounts to $\delta E^{\mu \nu} \delta_Y g_{\mu \nu} =0$. Since $\delta_{Y}g_{\mu\nu}\neq0$ in general, it follows $\delta E^{\mu\nu}=0$ in the Euclidean bulk $\mathcal{M}$.\footnote{Demanding $\delta E^{\mu \nu} \delta_Y g_{\mu \nu} =0$ for all Lorentzian initial data leads to $\delta E^{\mu\nu}=0$ everywhere in $\mathcal{M}$ \cite{Pedraza:2021mkh,Pedraza:2021fgp}.} Further, demanding (\ref{eq:domEIn}) hold in all Lorentz frames, one is able to conclude the \emph{Lorentzian} Einstein equations hold everywhere in the AdS cylinder.

The above derivation establishes, assuming CV duality, the first law of complexity implies the linearized Einstein's equations around vacuum AdS, or, more generally, another reference background where $|\lambda_i\rangle\neq|0\rangle$. The covariant derivation neatly encapsulates a notion of spacetime complexity: optimal quantum computation imposes gravitational field equations \cite{Pedraza:2022dqi}. To paraphrase Maupertuis, ``Nature is thrifty in all its \emph{computations}''.

The remainder of this article essentially follows \emph{mutatis mutandis}. By altering the form of the volume functional in complexity=volume (the bottom leg in Figure \ref{fig:trinity}), we will show how gravitational equations of motion for higher-order gravities, including semi-classical quantum corrections, arise. Importantly, we will maintain the boundary first law is the same across all theories, which is the case for the complexity defined in (\ref{eq:compftdef}) that is naturally suggested by holographic state preparation. In Section \ref{sec:disc} we ponder about possible extensions due to modifying other inputs in Figure \ref{fig:trinity}.

\section{Higher-derivative gravity from complexity} \label{sec:highordergravcomp}

Here we show how linearized equations of motion for higher-order theories of gravity arise from the first law of holographic complexity. Our inputs to the derivation include assuming the form of the boundary first law (\ref{eq:firstlawcomnograv}) and the first law of holographic complexity (\ref{eq:holofirstlawv1}). As in the case of black hole thermodynamics, where the horizon area is replaced by an area functional known as the Iyer-Wald entropy \cite{Wald:1993nt,Iyer:1994ys}, it is natural to assume geometric volume is to be replaced by a volume functional in the context of arbitrary diffeomorphism covariant theories of gravity. One proposal is the generalized volume $W_{\text{gen}}$  \cite{Bueno:2016gnv}
\begin{equation}\label{eq: generalizedVolume}
W_{\text{gen}}=\frac{1}{(d-2) P_{0}} \int_{\Sigma} \epsilon_{\Sigma}\left(P^{\mu\nu\rho\sigma} n_{\mu} n_{\sigma} h_{\nu\rho}-P_{0}\right).
\end{equation}
Here $\epsilon_{\Sigma}=\sqrt{h}d^{d-1}y$ is the induced volume form on the hypersurface $\Sigma$ with induced metric $h_{\mu\nu}$ and future-pointing timelike unit normal $n_{\mu}$, and $P^{\mu\nu\rho\sigma}$ is the Iyer-Wald tensor \cite{Iyer:1994ys}
\beq P^{\mu\nu\rho\sigma}\equiv \frac{\partial \mathcal{L}}{\partial R_{\mu\nu\rho\sigma}}-\nabla_{\mu_{1}}\frac{\partial\mathcal{L}}{\partial\nabla_{\mu_{1}}R_{\mu\nu\rho\sigma}}+...+(-1)^{m}\nabla_{(\mu_{1}}...\nabla_{\mu_{m})}\frac{\partial\mathcal{L}}{\partial \nabla_{(\mu_{1}}...\nabla_{\mu_{m})} R_{\mu\nu\rho\sigma}}\;,\label{eq:IWtensgen}\eeq
with $\mathcal{L}$ being the Lagrangian scalar density, $L=\mathcal{L}\epsilon$. In maximally symmetric spacetimes (MSS) the tensor will take the form $P^{\mu\nu\rho\sigma}_{\text{MSS}}=P_{0}(g^{\mu\rho}g^{\nu\sigma}-g^{\mu\sigma}g^{\nu\rho})$, with $P_{0}$ being a theory dependent constant. For example, in Einstein gravity, $\mathcal{L}=(16\pi G)^{-1}(R-2\Lambda)$ and $P^{\mu\nu\rho\sigma}=(32\pi G)^{-1}(g^{\mu\rho}g^{\nu\sigma}-g^{\mu\sigma}g^{\nu\rho})$ with $P_{0}\equiv(32\pi G)^{-1}$, recovering $W_{\text{gen}}=\int_{\Sigma}\epsilon_{\Sigma}=V$.

Another proposal for complexity in higher-derivative gravity \cite{Hernandez:2020nem} is that the generalized volume functional should be supplemented by corrections involving the extrinsic curvature, denoted $W_{K}$. Such a term is analogous to the `anomaly' contribution appearing in the Camps-Dong formula for holographic entanglement entropy for $F$(Riemann) theories of gravity \cite{Dong:2013qoa,Camps:2013zua}.
Then, the proposed CV prescription for higher-derivative theories (induced on holographic braneworlds) takes the form
\begin{equation}
\mathcal{C}(\sigma_{A})=\frac{1}{G\ell}\,\underset{\Sigma\sim A}{\text{max}}\left[W_{\mathrm{gen}}(\Sigma)+W_{K}(\Sigma)\right].
\label{eq:gencompprop}\end{equation}
Notably, the generalized volume $W_{\text{gen}}$ 
(\ref{eq: generalizedVolume}) takes a slightly different form than the one in (\ref{eq:gencompprop}) suggested by \cite{Hernandez:2020nem}.\footnote{Arguably, the $W_{\text{gen}}$ of \cite{Hernandez:2020nem} corresponds to a functional  proposed in \cite{Bueno:2016gnv} (cf. Eq. (66)) for a suitable choice of otherwise undetermined constants.} For our purposes these differences are unimportant, however, one could argue our analysis advocates for the functional (\ref{eq: generalizedVolume}) as it leads to the gravitational equations of motion.

Here, assuming only the boundary first law and replacing the standard CV formula (\ref{eq:CVcomplexitydef}) with the higher-derivative complexity formula (\ref{eq:gencompprop}), we show  
\begin{equation}\label{eq: firstLawHigherDerivative}
\delta_{\lambda_{f}}\mathcal{C}=\Omega_\text{bdry}( \delta_Y \tilde{\lambda}, \delta \tilde{\lambda})= \frac{1}{G\ell}\delta W_{\text{gen}}
\end{equation}
holds for linear perturbations around vacuum AdS for slices $\Sigma$ which have vanishing extrinsic curvature, including the constant $t$ surfaces considered in the previous section. Moreover, this holographic first law implies the bulk spacetime obey the linearized equations of motion for higher-derivative theories. 
We emphasize that, although in Einstein gravity the first law \eqref{eq: firstLawHigherDerivative} holds for variations around any spacetime satisfying Einstein's equations \cite{Pedraza:2021fgp,Pedraza:2021mkh}, this will not be the case for the higher-order theories. In particular, the first law \eqref{eq: firstLawHigherDerivative} is a special consequence of considering maximal slices in vacuum AdS which have vanishing extrinsic curvature since in this case both $W_K( \Sigma)$ and  $\delta W_K( \Sigma)$ vanish. This follows from the fact $W_K$ is thought to be quadratic in the extrinsic curvature \cite{Hernandez:2020nem}. It would be interesting to find the analog of $\Omega_\text{bdry}$ which includes the $W_K$ corrections, but we leave this for future work.

\subsection*{First law of holographic complexity in higher-order gravity}

Our main task is to establish that in higher-order gravity
\beq \Omega_{\text{bdry}}(\delta_{Y}\tilde{\lambda},\delta\tilde{\lambda})= \frac{1}{G\ell}\delta W_{\text{gen}}\;,\label{eq: firstThingToProveHigherDerivative}\eeq
where $\Sigma$ is a maximal slice with vanishing extrinsic curvature $K_{ab}$. Indeed, the equivalence of boundary and bulk symplectic forms (\ref{eq: boundaryEqualsInitial}) holds for arbitrary theories of gravity (assuming the linearized equations of motion). Thus, here we confirm $\Omega_{\text{bulk}}(g,\delta_{Y}g,\delta g)=\delta W_{\text{gen}}$, at least for perturbations about vacuum AdS.  

To this end, let us start by considering the bulk behavior of the new York transformation (\ref{eq: yorkTransformation}) about vacuum AdS, a MSS spacetime (see \cite{Belin:2018bpg} and Appendix \ref{app:NYVaccumAdS}). In Euclidean signature, the York deformation is implemented by the spacetime vector field $Y\to \equiv iY_{E}$ for Euclidean time $\tau_{E}$. Generically, 
\begin{equation}
    \delta_Y g_{\mu\nu}=\mathcal{L}_{Y}g_{\mu\nu}=\beta(\tau_{E})h_{\mu\nu}\;,
\label{eq:Yorkdefgencoord}\end{equation}
where $h_{\mu\nu}=g_{\mu\nu}+n_{\mu}n_{\nu}$ is the induced metric on $\Sigma$, and we used the fact that in vacuum AdS the new York transformation acts as a diffeomorphism. In particular, in Wheeler-de Witt (WdW) coordinates, $Y=i\alpha \partial_{\tau_{E}}$ for some real parameter $\alpha$
\begin{equation}
    ds^2_{d}=g_{\mu\nu}(x)dx^{\mu}dx^{\nu}=d\tau_{E}^2+h_{ab}(\tau_{E},y)dy^{a}dy^{b}\;,
\end{equation}
with $h_{ab}(\tau_{E},y)=\cosh^{2}(\tau_{E})\sigma_{ab}(y)$ (being the induced metric projected onto surfaces of constant $\tau_{E}$, $h_{ab}=e^{\mu}_{a}e^{\nu}_{b}h_{\mu\nu}$) and $\sigma_{ab}$ the induced metric on $\Sigma$ at $\tau_E=0$. Then, one finds
\beq \mathcal{L}_{Y}h_{ab}=i\alpha\partial_{\tau_{E}}h_{ab}=2i\alpha \tanh(\tau_{E})h_{ab}\;,\eeq
and $\mathcal{L}_{Y}g_{\tau_{E} \tau_{E}}=\mathcal{L}_{Y}g_{\tau_{E}a}=0$, obeying (\ref{eq:Yorkdefgencoord}). Notice that the function $\beta(\tau_{E})=2i\alpha \tanh\tau_{E}$ vanishes at $\tau_{E}=0$, the maximal Cauchy surface $\Sigma$ with vanishing $K$. Thus, instantaneously at $\tau_{E}=0$, the vector field $Y$ is Killing in that it obeys Killing's equation. In Poincar\'e coordinates one also finds $\mathcal{L}_{Y}g_{\mu\nu}|_{\Sigma}=\beta(t)h_{\mu\nu}$ (see Appendix \ref{app:NYVaccumAdS}).

 Moreover, since $\beta$ vanishes on $\Sigma$, its gradient will be proportional to the normal $n_{\mu}$ on $\Sigma$
 \begin{equation}
    \nabla_{\sigma}(\delta_Y g_{\mu\nu})|_{\Sigma}=\nabla_{\sigma}(\beta h_{\mu\nu})|_\Sigma=\frac{1}{N}n_{\sigma} h_{\mu\nu},
\label{eq:nabofdYgen}\end{equation}
where $N$ is a constant over $\Sigma$, and we used $(\beta\nabla_{\sigma}h_{\mu\nu})|_{\Sigma}=0$. Specifically, in WdW coordinates, surfaces of constant $\tau_{E}$ have unit normal $n_{\mu}=i(\partial_{\tau_{E}})_{\mu}$, meanwhile 
\beq \nabla_{\mu}\beta=\partial_{\mu}\beta=\frac{2i\alpha}{\cosh^{2}(\tau_{E})}\delta^{\tau_{E}}_{\mu}\;,\eeq
such that at $\tau_{E}=0$, $\nabla\beta=2\alpha i\partial_{\tau_{E}}$. Hence, we recover (\ref{eq:nabofdYgen}), identifying $N=1/2\alpha$ and $n_{\mu}=N\nabla_{\mu}\beta$. 

It is worth pausing briefly to compare to the geometric set-up employed in \cite{Jacobson:2015hqa,Bueno:2016gnv}, where gravitational equations of motion were derived via the `entanglement equilibrium' hypothesis. There, one compares surface areas of small spatial balls $\Sigma$ in maximally symmetric spacetimes to those in spacetimes that are perturbations away from a MSS. The causal diamond of $\Sigma$ in a MSS, defined as the union of past and future domains of dependence of $\Sigma$, is generated by a conformal Killing vector field $\zeta^{\mu}$ that preserves the diamond. As a conformal Killing vector, $\zeta$ obeys the conformal Killing equation, $\mathcal{L}_{\zeta}g_{\mu\nu}=2\beta g_{\mu\nu}$, where conformal factor $\beta=\frac{1}{d}\nabla_{\mu}\zeta^{\mu}$ vanishes at $\Sigma$. Additionally, the gradient of $\beta$ is proportional to the unit normal on $\Sigma$, as above, for constant $N=(d-2)/k\kappa$, where $k$ is the trace of the extrinsic curvature of $\partial\Sigma$ embedded in $\Sigma$ and $\kappa$ is the surface gravity of the conformal Killing horizon. Moreover, $\nabla_{\mu}(\mathcal{L}_{\zeta}g_{\alpha\beta})|_{\Sigma}=\frac{2}{N}n_{\mu}g_{\alpha\beta}$.

Thus, 
the new York transformation and the conformal Killing flow appear to be related, despite the vector field $Y$ not being a conformal Killing vector. There is in fact a relationship between the two, as noted in \cite{Jacobson:2018ahi}. Namely, the conformal Killing transformation and new York transformation are equivalent \emph{only} on the maximal hypersurface $\Sigma$ of a causal diamond, where $\mathcal{L}_{\zeta}h_{\mu\nu}|_{\Sigma}=0$ and $\mathcal{L}_{\zeta}K_{\mu\nu}|_{\Sigma}=-\alpha h_{\mu\nu}$, upon identifying $\alpha=-n^{\mu}\nabla_{\mu}\beta_{\text{CD}}|_{s=0}$, where $s$ is conformal Killing time ($s=0$ coincides with $\Sigma$). The relation between the diamond preserving conformal Killing flow and new York deformation is not so surprising since hypersurfaces of constant $s$ are also slices of constant mean curvature $K$, where $K=0$ when $s=0$. In this way, conformal Killing time is akin to York time. 

Importantly, the similarities between the new York transformation and conformal Killing flow preserving a causal diamond allow us to derive precisely the same generalized volume functional (\ref{eq: generalizedVolume}) found in the context of the first law of causal diamond mechanics \cite{Bueno:2016gnv}. Our calculation follows nearly identical to \cite{Bueno:2016gnv}, with only a handful of differences. We summarize the key formulae, relegating additional computational details for Appendix \ref{app:generalized volume}.

Consider a higher-order theory of pure gravity with Lagrangian $d$-form of the type
\beq
    L=L(g_{\mu\nu},R_{\mu\nu\rho\sigma},\nabla_{\mu_1} R_{\mu\nu\rho\sigma},...,\nabla_{(\mu_1}...\nabla_{\mu_m)}R_{\mu\nu\rho\sigma}).
\eeq
For such a theory,  the symplectic current $(d-1)$ form can be written as (see \cite{Iyer:1994ys} for a proof)
\begin{equation}
\begin{split}
        \omega_{\text{bulk}}(g,\delta_1 g,\delta_2 g)=&2 \delta_1 P^{\nu\rho\sigma}\nabla_{\sigma}\delta_2 g_{\nu\rho}-2P^{\nu\rho\sigma}\delta_1 \Gamma_{\phantom{e}\nu\sigma}^{\gamma}\delta_2g_{\gamma\rho}+\delta_1 S^{\mu\nu}\delta_2g_{\mu\nu}\\
    &+\displaystyle\sum_{i=1}^{m-1}\delta_1 T_{i}^{\mu\nu\rho\sigma\mu_1...\mu_i}\delta_2\nabla_{(\mu_1}...\nabla_{\mu_i)} R_{\mu\nu\rho\sigma}-(1\leftrightarrow 2),
\end{split}
\label{eq: higherDerivativeSymplecticFormn}
\end{equation}
with $P^{\nu\rho\sigma}=\epsilon_\mu P^{\mu\nu\rho\sigma}$, $S^{\mu\nu}$ and $T_i^{\mu\nu\rho\sigma \mu_1...\mu_i}$ are functions of the metric, the Riemann tensor and its covariant derivatives. We are interested in deformations about empty AdS, a MSS. For such spacetimes, the Riemann tensor takes the form $R_{\mu\nu\rho\sigma}=R(g_{\mu\rho}g_{\nu\sigma}-g_{\mu\sigma}g_{\nu\rho})/d(d-1)$ with constant $R$. Therefore, $\nabla_{\gamma}R_{\mu\nu\rho\sigma}=0$, and $\mathcal{L}_{Y}R_{\mu\nu\rho\sigma}=0$ on the maximal hypersurface $\Sigma$. Likewise, the tensors $P^{\mu\nu\rho\sigma}$, $S^{\mu\nu}$ and $T_i^{\mu\nu\rho\sigma \mu_1...\mu_i}$ will have vanishing Lie derivative along $Y$ evaluated on $\Sigma$. Then, evaluating (\ref{eq: higherDerivativeSymplecticFormn}) with the new York variation, we find
\begin{equation}
\begin{split}
    \omega_{\text{bulk}}(g,\delta_Y g,\delta g)|_\Sigma=
    -\frac{2}{N}\left[2n_{\sigma} h_{\nu\rho}\delta P^{\nu\rho\sigma}+P^{\nu\rho\sigma}(n_\nu h_{\sigma}^\gamma+n_{\sigma}h_{\nu}^\gamma-n^\gamma h_{\nu\sigma})\delta g_{\gamma\rho}\right]\;,
\end{split}
\end{equation}
where we used (\ref{eq:nabofdYgen}).\footnote{Comparing to Eq. (29) of \cite{Bueno:2016gnv}, there is an overall sign difference since their they compute $\omega(g,\delta g,\mathcal{L}_{\zeta}g)$.}

 We would like to rewrite the symplectic current as a variation of some scalar functional. Taking inspiration from \cite{Bueno:2016gnv}, this is accomplished by introducing a tensor $F^{\mu\nu\rho\sigma}\equiv P^{\mu\nu\rho\sigma}-P_{0}(g^{\mu\rho}g^{\nu\sigma}-g^{\mu\sigma}g^{\nu\rho})$, characterizing the difference between $P^{\mu\nu\rho\sigma}$ and its background value. Clearly $F^{\mu\nu\rho\sigma}$ vanishes identically in empty AdS. Consequently, substituting $F^{\mu\nu\rho\sigma}$  for $P^{\mu\nu\rho\sigma}$ results in
 \begin{equation}\label{eq: omegafinal}
    \omega_{\text{bulk}}(g,\delta_Y g,\delta g)|_{\Sigma}=\frac{2}{N}\delta\left[\epsilon_{\Sigma}\left(P^{\mu\nu\rho\sigma}n_\mu n_\sigma h_{\nu\rho}-P_0\right)\right].
\end{equation}
 where we used that any term dependent on $\delta F^{\mu\nu\rho\sigma}$ may be replaced by a total variation since variations of other tensors will be multiplied by the background value of $F^{\mu\nu\rho\sigma}$ which is identically zero.  Since $N$ is constant, the bulk symplectic form is
\beq \Omega_{\textup{bulk}}(g,\delta_Y g ,\delta g)=\int_{\Sigma}\omega_{\text{bulk}}(g,\delta_{Y}g,\delta g)=4P_0(d-2)\alpha\delta W_{\textup{gen}}=\frac{1}{G\ell}\delta W_{\text{gen}}\;,
\eeq
with generalized volume (\ref{eq: generalizedVolume}), and we set parameter $\alpha\equiv[4(d-2)P_{0}G\ell]$. In the case of Einstein gravity, we recover the bulk relationship (\ref{eq:OmbulkdeltaV}) with $\alpha=8\pi/(d-2)\ell$.  In \cite{Bueno:2016gnv}, the variation of the generalized volume is proportional to the variation of the gravitational Hamiltonian $\delta H_{\zeta}=\int_{\Sigma}\omega(g,\delta g,\mathcal{L}_{\zeta}g)$. This suggests the generalized volume $W_{\text{gen}}$ should be interpreted as a Hamiltonian. Indeed, in general relativity, the volume $V=\int_{\Sigma}\epsilon_{\Sigma}$ is understood as a Hamiltonian in the CMC slicing \cite{York:1972sj,Belin:2018bpg}. 

Having established that, for deformations about vacuum AdS, the bulk symplectic form is proportional to the variation of the generalized volume, we can reverse engineer the steps leading to the equivalence between boundary and bulk symplectic forms to yield
\beq \Omega_{\text{bdry}}(\delta_{Y}\tilde{\lambda},\delta\tilde{\lambda})=\Omega_{\text{bulk}}(g,\delta_{Y}g,\delta g)=\frac{1}{G\ell}\delta W_{\text{gen}}\;.\label{eq:bdrybulksympformhigh}\eeq
We emphasize this is only true when both $\delta g$ and $\delta_{Y}g$ obey the linearized equations of motion, i.e., $\delta E_{\mu\nu}=0$ and $\delta_{Y}E_{\mu\nu}=0$. The first of these statements is an assumption, while the second is only true for deformations about vacuum AdS, the case under consideration here, since the new York deformation acts as a diffeomorphism. To extend the relation (\ref{eq:bdrybulksympformhigh}) to more general states requires knowledge of how the York transformation in other theories. Indeed, even in Einstein gravity, the new York deformation is not a diffeomorphism in general. Nonetheless, upon solving the constraints of the theory one has $\Omega_{\text{bdry}}(\delta_{Y}\tilde{\lambda},\delta\tilde{\lambda})=\Omega_{\text{bulk}}(g,\delta_{Y}g,\delta g)$ only on the maximal slice $\Sigma$ where the trace of the extrinsic curvature vanishes, $K=0$. For higher-order gravities, we expect, when the constraints are solvable, that the new York deformation would be on-shell for a more complicated constraint on the extrinsic curvatures of the slice $\Sigma$. It would be interesting to see whether extremizing the more general volume functional (\ref{eq:gencompprop}) results in the same condition on the extrinsic curvatures. That is, whether the condition of the new York deformation being on-shell is consistent with the extremization condition of the generalized volume functional. This is akin to what happens in holographic entanglement entropy in higher-curvature theories of gravity, where the surface which extremizes the Camps-Dong entropy functional may not be consistent with the bulk equations of motion \cite{Bhattacharyya:2014yga,Erdmenger:2014tba}. 

Lastly, since we continue to assume the same boundary first law (\ref{eq:firstlawcomnograv}), the holographic first law of complexity for CFTs dual to higher-order theories of gravity is as presented in (\ref{eq: firstLawHigherDerivative}).
This is tantamount to modifying the bottom leg of the triangle in Figure \ref{fig:trinity}.

\subsection*{Higher-order equations of motion from the first law}

Having established that the bulk symplectic form $\Omega_{\text{bulk}}(g,\delta_{Y}g,\delta g)$ is proportional to the variation of the generalized volume $W_{\text{gen}}$, the derivation of the linearized equations of motion for arbitrary higher-order theories of gravity follows in precisely the same way as in Einstein gravity reviewed in Section \ref{sec:firstlawholocomp}. All that is required is to assume the boundary and holographic first laws such that 
\beq \frac{1}{G\ell}\delta W_{\text{gen}}=\Omega_{\text{bdry}}(\delta_{Y}\tilde{\lambda},\delta\tilde{\lambda})\Rightarrow \delta E_{\mu\nu}=0\;,\eeq
where $\delta E_{\mu\nu}=0$ now denotes the linearized equations for any higher-order theory of gravity (for perturbations about vacuum AdS), extending (\ref{eq:firstlawassumption}). As in the Einstein case, $d\omega_{\text{bulk}}^{E}=0$ when $\delta E_{\mu\nu}=0$. Thus, for perturbative excited states with $\mathcal{O}(N^0)$ backreaction, linearized gravitational dynamics for \emph{any} theory of gravity emerges from optimized computation.

\section{Semi-classical gravity from complexity} \label{sec:semiclassgravcomp}

We have seen how the linearized equations of motion for any classical diffeomorphism invariant theory of gravity follow from the first law of holographic complexity, under a suitable modification to the volume functional appearing in complexity-volume. Here we see how the situation changes in the presence of quantum corrections due to semi-classical backreaction. In spacetime dimensions $d\geq4$, the problem of backreaction is notoriously difficult, as it requires solving the semi-classical Einstein equations in its regime of validity.  Notably, however, the problem of backreaction is exactly solvable in models of two-dimensional dilaton gravity, including semi-classical Jackiw-Teitelboim (JT) gravity \cite{Jackiw:1984je,Teitelboim:1983ux}. This is because, up to relatively minor ambiguities, backreaction effects are largely fixed by the Polyakov action capturing contributions of the two-dimensional conformal anomaly \cite{Polyakov:1981rd}. We therefore analyze CV complexity and the new York transformation in semi-classical JT gravity. Previous studies of holographic complexity in classical two-dimensional dilaton models have been presented in, e.g., \cite{Brown:2018bms,Goto:2018iay,Chapman:2021eyy,Anegawa:2023wrk,Bhattacharya:2023drv}. Semi-classical corrections were considered in \cite{Schneiderbauer:2019anh,Schneiderbauer:2020isp} in the context of the soluble Russo-Thorlacius-Susskind (RST) model \cite{Russo:1992ax,Russo:1992ht}.

\subsection{Jackiw-Teitelboim gravity: a 2D case study}\label{sec:holocomJT}

\subsubsection{Classical analysis}

Before we include semi-classical backreaction effects, consider classical JT gravity characterized by the action, with a Gibbons-Hawking-York (GHY) boundary term and a local counterterm
\begin{equation}
I_{\mathrm{JT}}=I_{\mathrm{JT}}^{\mathrm{bulk}}+I_{\mathrm{JT}}^{\mathrm{GHY}}+I_{\mathrm{JT}}^{\mathrm{ct}},
\label{eq:JTact1}\end{equation}
\begin{align} 
    & I_{\mathrm{JT}}^{\mathrm{bulk}}=\frac{1}{16 \pi G_{2}} \int_{\tilde{\mathcal{M}}} d^2 x \sqrt{-g}\left(\left(\Phi_0+\Phi\right) R+\frac{2 \Phi}{L^2}\right),\\ & I_{\mathrm{JT}}^{\mathrm{GHY}}+I_{\mathrm{JT}}^{\mathrm{ct}}=\frac{1}{8 \pi G_{2}} \int_\mathcal{B} d t \sqrt{-\gamma}\left(\left(\Phi_0+\Phi\right) \mathcal{K}-\frac{\Phi}{L}\right) .
\end{align}
Here $G_{2}$ refers to a dimensionless two-dimensional Newton's constant,\footnote{There is no intrinsic notion of a two-dimensional Newton's constant. Rather, the prefactor $(\Phi_{0}+\Phi)$ plays the role a Newton's constant, where the dilaton diverges at the conformal boundary, indicating a region of weak gravity. We keep $G_{2}$ as a bookkeeping device.} $\tilde{\mathcal{M}}$ is the Lorentzian spacetime manifold with timelike boundary $\mathcal{B}$, $\Phi$ is the dilaton arising from a spherical reduction of the parent theory, $\Phi_0$ is a constant proportional whose physical significance will be commented on momentarily, $L$ is the $\mathrm{AdS}_2$ length scale which we will subsequently set to unity, and $\mathcal{K}$ is the trace of the extrinsic curvature of $B$ with induced metric $\gamma_{\mu \nu}$. The gravitational and dilaton equations of motion are, respectively, 
\beq T_{\mu\nu}\equiv-\frac{2}{\sqrt{-g}}\frac{\delta I_{\text{JT}}}{\delta g^{\mu\nu}}=-\frac{1}{8\pi G_{2}}\left(g_{\mu\nu}\Box-\nabla_{\mu}\nabla_{\nu}-g_{\mu\nu}\right)\Phi=0\;,\label{eq:gravEOMclassJT}\eeq
\beq R+2=0\;.\label{eq:dilaEOMJT}\eeq
Thus, the dilaton equation of motion fixes the background to be empty $\text{AdS}_{2}$. In Poincar\'e coordinates,
\begin{equation}
    ds^2 = \frac{1}{z^2}(-dt^2 + dz^2),
\end{equation}
the gravitational equations of motion (\ref{eq:gravEOMclassJT}) admit the linearly varying dilaton 
\beq \Phi=\frac{1}{2\mathcal{J}z}\;,\eeq
as one of its solutions, with $\mathcal{J}$ being an energy scale which characterizes how the $SL(2,\mathbb{R})$ isometries of $\text{AdS}_{2}$ are broken to $U(1)$ under linear variations in $\Phi$ \cite{Maldacena:2016upp}. We see $\Phi$ diverges at the conformal boundary, $z=0$.

One perspective of the JT action is that the theory characterizes the physics of higher-dimensional near-extremal black holes in the near-horizon limit. The action may be derived via a spherical dimensional reduction where the dilaton $\Phi$ controls the size of the sphere, and the constant $\Phi_{0}$ is proportional to the extremal entropy of the higher-dimensional black hole, where $\Phi_{0}\gg\Phi$ such that $\Phi$ encodes deviations from extremality. Another perspective is that classical JT gravity may be interpreted as the gravitational dual to the Sachdev-Ye-Kitaev (SYK) model of interacting fermions \cite{Sachdev:1992fk,Kit_SYK}. There are a  plethora of studies on operator growth and complexity in SYK, e.g., \cite{Roberts:2018mnp,Qi:2018bje,Parker:2018yvk,Barbon:2019wsy,Jian:2020qpp}, however, we will be agnostic to the precise microscopic dual to JT gravity. Rather, we treat JT gravity and its semi-classical extension below as an effective toy model to study the problem of backreaction.

Our goal in this section is to first determine the analog of the new York deformation $\delta_{Y}$ such that the bulk symplectic form is proportional to the variation of the volume in classical JT gravity. Second, we will show that assuming the first law of holographic complexity and the same boundary first law is enough to imply the linearized JT equations of motion.

\subsection*{Volume and JT Hamiltonian}

CV complexity in JT gravity was previously studied in \cite{Brown:2018bms}, where complexity is assumed to be dual to the volume of the extremal slice $\Sigma$ with an important proportionality factor
\beq \label{eq: jtvolume1}
V_{\text{JT}} \sim \Phi_0\int_{\Sigma} dy \sqrt{h}\;.
\eeq
This choice is motivated by the fact complexity is expected to grow at a rate
 proportional to the number of degrees of freedom of the dual quantum system. Viewing JT gravity as the effective dynamics of a near-extremal black hole, the number of degrees of freedom is proportional to the black hole entropy, $S_{\text{BH}}\sim(\Phi_{0}+\Phi)$, with $\Phi_{0}\gg\Phi$. However, as indicated in \cite{Anegawa:2023wrk,Bhattacharya:2023drv}, including the dilaton $\Phi$ leads to non-trivial subleading corrections. Thus, we take the volume functional to be (setting $16\pi G_{2}=1$)
\beq \label{eq: jtvolume}
V_{\text{JT}} = \int_{\Sigma} dy \ \sqrt{h}(\Phi+\Phi_0)\;,
\eeq
consistent with the volume functional considered in flat models of dilaton gravity \cite{Schneiderbauer:2019anh}.\footnote{Note that this form of the volume functional does not directly follow from the generalized volume (\ref{eq: generalizedVolume}) for higher-order theories, in contrast for what happens in the case of black hole entropy \cite{Iyer:1994ys}. The derivation of the generalized volume must be appended to account for non-minimally coupled dilaton theories of gravity.} Note that this means, for $\text{AdS}_{2}$ black holes, the complexity=volume prescription requires one to extremize this functional, which is not equivalent to calculating a two-dimensional geodesic length, as performed in \cite{Brown:2018bms}. Additionally, $\Sigma$ is not a surface where $K=0$ (except for in the limit $\Phi=0$), due to the fact the dilaton is a function on spacetime. In particular, using the definition of the extrinsic curvature tensor, the condition for $\Sigma$ to extremize \eqref{eq: jtvolume} is
\beq\label{eq:extremalJT}
  K (\Phi+\Phi_0) + \frac{1}{N}(\dot{\Phi} - N^a \nabla_a \Phi) =0\;, 
\eeq
which is modified due to the dilaton term. When $\Phi=1$ and $\Phi_{0}=0$, this reduces to the familiar $K=0$ condition. 

We are interested in the analog of the new York deformation such that the bulk symplectic form is proportional to the variation of the volume (\ref{eq: jtvolume}), and preserves the constraints of JT gravity. To this end,  we consider JT gravity in the ADM formalism (see Appendix \ref{app:ADMforms}). Note that in two dimensions $K_{ab}=Kh_{ab}$, and, subsequently, $\pi^{ab}=\pi h^{ab}$. Specifically, we find
\beq
    K^{ab} = - \frac{h^{ab} \pi_{\Phi}}{2 \sqrt{h}}\;,\quad \dot{\Phi} = -\frac{N \pi}{\sqrt{h}} + N^a \nabla_a \Phi\;,
\label{eq: dotphi}\eeq
where the extrinsic curvature tensor is defined via $\dot{h}_{ab} = 2N K_{ab} + \nabla_a N_b + \nabla_b N_a$. The Hamiltonian may be cast, up to a boundary term unimportant to us, as
\beq H_{\text{JT}}=\int_{\Sigma_{t}}dy\sqrt{h}(N\mathcal{H}_{\text{JT}}+N_{a}\mathcal{H}_{\text{JT}}^{a})\;,\eeq
with Hamiltonian and momentum constraints, respectively,
\beq
\begin{split}
&\mathcal{H}_{\text{JT}}=\frac{2\pi K}{\sqrt{h}}-(\Phi_{0}+\Phi)\bar{R}+2\Box\Phi-2\Phi\;,\\
&\mathcal{H}^{a}_{\text{JT}}=\frac{\pi_{\Phi}}{\sqrt{h}}\nabla^{a}\Phi-2\nabla_{b}\left(\frac{\pi h^{ab}}{\sqrt{h}}\right)\;,
\end{split}
\label{eq:momconstraintsJT}\eeq
where $\bar{R}$ is the intrinsic Ricci scalar. It can be shown these constraints are first class and $\mathcal{H}^{a}_{\text{JT}}$ is the generator of spatial diffeomorphisms on $\Sigma_{t}$ while $\mathcal{H}_{\text{JT}}$ generates time translations. Moreover, the constraints are simple enough that they can be solved explicitly (cf. \cite{Henneaux:1985nw,Louis-Martinez:1993bge,Iliesiu:2020zld}).

From our ADM split we can determine the bulk symplectic form $\Omega_{\text{bulk}}$ with respect to phase space variables $(h_{ab},\pi^{ab},\Phi,\pi_{\Phi})$. To see this, note that the ADM action may be cast as 
\beq I_{\text{ADM}}=\int dt\int_{\Sigma_{t}}dy(\pi^{ab}\dot{h}_{ab}+\pi_{\Phi}\dot{\Phi}-\mathcal{H}_{\text{ADM}})+I_{\mathcal{B}}\;,\eeq
where the term in parentheses is the ADM Lagrangian and $I_{\mathcal{B}}$ is some appropriate boundary term.  Varying the action we find
\beq \delta I_{\text{ADM}}=\int_{\Sigma_{t}}dy(\pi^{ab}\delta h_{ab}+\pi_{\Phi}\delta\Phi)+\text{bulk EOMs}\;.\eeq
Comparing to the standard variation of the action
\beq \delta I_{tot}=\int_{\tilde{\mathcal{M}}}E_{\phi}\delta\phi+\int_{\Sigma}\theta(\phi,\delta\phi)\;,\eeq
where $\phi=\{h_{ab},\Phi\}$ and $E_{\phi}$ represents the equation of motion form for each field (with an implicit sum over field type $\phi$),
we are able to read off the symplectic potential $\theta$
\beq \theta\equiv\theta_{h}+\theta_{\Phi}\;,\quad \theta_{h}=\pi^{ab}\delta h_{ab}\;,\quad \theta_{\Phi}=\pi_{\Phi}\delta\Phi\;.\eeq
Further, the symplectic current form similarly decomposes as  $\omega=\omega_{h}+\omega_{\Phi}$
 with
 \beq \omega_{h}=\delta_{1}\pi^{ab}\delta_{2}h_{ab}-\delta_{2}\pi^{ab}\delta_{1}h_{ab}\;,\quad \omega_{\Phi}=\delta_{1}\pi_{\Phi}\delta_{2}\Phi-\delta_{2}\pi_{\Phi}\delta_{1}\Phi\;.\eeq
Therefore, the bulk symplectic form for classical JT gravity is
\beq \Omega_{\text{JT}}(\phi,\delta_{1}\phi,\delta_{2}\phi)=\int_{\Sigma_{t}}(\delta_{1}\pi^{ab}\delta_{2}h_{ab}-\delta_{2}\pi^{ab}\delta_{1}h_{ab})+\int_{\Sigma_{t}}(\delta_{1}\pi_{\Phi}\delta_{2}\Phi-\delta_{2}\pi_{\Phi}\delta_{1}\Phi)\;.\label{eq:bulksympformclassJT}\eeq

\subsection*{A JT `new York' deformation}

Our goal now is to determine the analog of the new York deformation in JT gravity, i.e., the variation $\delta_{Y}$ such that 
\begin{equation}\label{eq:JTVol}
     \Omega_{\text{JT}}(\phi,\delta_Y \phi, \delta \phi) = \alpha \delta V_{JT} = \alpha \int_{\Sigma_{t}}dy[(\delta{\sqrt{h}})(\Phi + \Phi_0) + \sqrt{h} \delta \Phi].
\end{equation}
It is easy to verify the deformation which achieves \eqref{eq:JTVol} satisfies
\beq
\delta_Y \pi^{ab} = \frac{\alpha}{2}\sqrt{h} h^{ab} (\Phi + \Phi_0), \quad \delta_Y \pi_{\Phi} = \alpha \sqrt{h},\quad \delta_{Y}h_{ab}=\delta_{Y}\Phi=0\;.
\label{eq:nuNYtransJT}\eeq
In terms of configuration space variables \eqref{eq: dotphi}, the new York deformation behaves as 
\beq \label{eq: configvariables}
\delta_Y K = - \frac{\alpha}{2},  \quad  \delta_Y \dot{\Phi} = - \frac {\alpha N} {2}(\Phi+ \Phi_0).
\eeq
Moreover, the deformation preserves the constraints of the theory (\ref{eq:momconstraintsJT}), 
\beq \delta_{Y}\mathcal{H}^{a}_{\text{JT}}=(\delta_{Y}\pi_{\Phi})\nabla^{a}\Phi-2\sqrt{h}\nabla^{a}\left(\frac{h_{cd}\delta_{Y}\pi^{cd}}{\sqrt{h}}\right)=0\;,\eeq
and 
\beq \delta_{Y}\mathcal{H}_{\text{JT}}=\frac{2}{\sqrt{h}}\delta_{Y}(\pi K)=\alpha\left[K(\Phi_{0}+\Phi)+\frac{1}{N}(\dot{\Phi}-N^{a}\nabla_{a}\Phi)\right]\;.\eeq
For $\delta_{Y}\mathcal{H}_{\text{JT}}=0$, we require the term in brackets to vanish. This is in fact precisely the condition the slice $\Sigma$ extremizes the volume $V_{\text{JT}}$, \emph{viz.}, \eqref{eq:extremalJT}.

As in higher dimensions, around the vacuum, the new York transformation acts as a diffeomorphism. To see this, consider the action of the two diffeomorphisms generated by the vector fields (in Poincar\'e coordinates for convenience)
\beq
\xi_0=\frac{\alpha z^{2}}{2\sqrt{z^{2}-t^{2}}}\partial_{t}+\frac{\alpha zt}{2\sqrt{z^{2}-t^{2}}}\partial_{z}, \quad \xi_1 = \alpha tz \mathcal{J} \Phi_0 \partial_z +  \frac{ \alpha \mathcal{J} \Phi_0}{2} (t^2 + z^2) \partial_t \;,
\eeq
where respective field variations are denoted by $\delta_0 \phi = \mathcal{L}_{\xi_0} \phi$ and $\delta_1 \phi = \mathcal{L}_{\xi_1} \phi$. The vector field $\xi_0$ is the usual spacetime vector field generator (see \ref{eq:YspacevecfieldGR}) of the new York transformation for Einstein gravity.  On the metric variables, this acts to enforce $\delta_0 h_{ab}=0$ and  $\delta_0 K = -\frac{\alpha}{2}$. Meanwhile, $\delta_0 \Phi|_{t=0}=0$ and
\beq 
\delta_0 \dot{\Phi}|_{t=0} = - \frac{\alpha }{2 z} \Phi = - \frac{\alpha N }{2} \Phi \;, 
\eeq
where we used $N=1/z$. Thus, the usual new York transformation of Einstein gravity gives the $\Phi_0$ independent contribution to the new York deformation of JT gravity defined in \eqref{eq:nuNYtransJT}. The second vector field, $\xi_{1}$, is a special conformal transformation and Killing vector in AdS.
 Since it is a Killing vector, the metric and extrinsic curvature variables remain unchanged under the diffeomorphism generated by $\xi_1$. On the other hand, the dilaton potential changes according to
\begin{equation}
    \delta_{1}\Phi= - \alpha \Phi_0 \frac{t}{2 z}, \quad \delta_1\dot{\Phi} = - \alpha \Phi_0 \frac{N}{2}\;.
\end{equation}
$\delta_{1}\Phi|_{t=0}=0$ and so we find the remaining $\Phi_0$ contribution to the deformation in \eqref{eq:nuNYtransJT}. 

The total transformation \eqref{eq:nuNYtransJT} is thus described by the vector field
\begin{equation}
     Y=  \xi_0 + \xi_1 \; ,
\label{eq:totalY}\end{equation}
which defines the deformation $\delta_Y$ through $\delta_Y \phi = \mathcal{L}_{Y} \phi $. Note that the new York transformation is not unique on the space of solutions to the dynamics since one can always act with a local diffeomorphism without changing the initial data on $\Sigma$ to obtain a physically equivalent solution. This is represented by a freedom in the choice of $Y$. In particular, we can apply local diffeomorphisms generated by $\xi^{\mu} \partial_{\mu}$ where $\xi^{\mu}|_{t=0} = 0, \partial_{\nu} \xi^{\mu}|_{t=0} =0$. For example, we can expand $\xi^{\mu} = \sum_{i=2} f^{\mu}_i(z) t^i$ and we still find agreement with the new York transformation on the surface $t=0$. 

\subsection*{A holographic first law of complexity and equations of motion}

In what follows we assume the same boundary first law, i.e., $\delta_{\lambda_{f}}\mathcal{C}=\Omega_{\text{bdry}}$. This amounts to assuming the dual quantum mechanical theory to JT gravity has a path integral representation which allows us to describe coherent states of the dual theory in terms of sources. Then, using the York deformation specialized to JT gravity (\ref{eq:nuNYtransJT}), we have the following first law of holographic complexity
\beq \delta_{\lambda_{f}}\mathcal{C}=\Omega_{\text{bdry}}(\delta_{Y}\tilde{\lambda},\delta\tilde{\lambda})=\frac{1}{G\ell}\delta V_{\text{JT}}\;.\label{eq:firstlawJT}\eeq
The second equality follows for linear perturbations around vacuum AdS (such that $\delta_{Y}$ is a diffeomorphism), and for slices $\Sigma$ obeying the extremization condition \eqref{eq:extremalJT}, such that the linearized equations of motion of the metric and dilaton hold, $\delta E_{\mu\nu}^{g}=\delta E_{\Phi}=0$.

As before, we can reverse the order in logic to derive the linearized equations of motion from the first law of complexity (\ref{eq:firstlawJT}). Namely, recall the steps leading to \eqref{eq: pushingVolumeToBoundary}, where now
\begin{equation}
\begin{split} \label{eq: JTfirstlaw}
i\int_{ \mathcal{M}_{-}} d \omega^{E}_{\text{JT}}( \phi,\delta_Y \phi, \delta \phi) & =\Omega_{\text{bdry}}( \delta_Y \tilde{\lambda}, \delta\tilde{\lambda}) - \frac{1}{G \ell}\delta V_{JT}
\end{split}
\end{equation}
where we have chosen $\alpha = \frac{1}{G \ell}$. 
In line with the higher-dimensional theories of gravity, the holographic first law is equivalent to the condition that the linearized equations of motion hold in the bulk. Specifically, when the first law (\ref{eq:firstlawJT}) holds we have $\int_{\mathcal{M}_{-}}d\omega_{\text{bulk}}^{E}=0$, i.e., 
\beq \label{eq:preeomvacuumJT}
\int_{\mathcal{M}_{-}}[\delta E^{\mu \nu}_g \delta_Y g_{\mu \nu} + \delta E_{\Phi}\delta_Y \Phi] =0 \;,
\eeq
for deformations around the vacuum, where $\delta_{Y}$ acts a diffeomorphism. Since the new York deformation is only defined up to local spacetime diffeomorphisms, (\ref{eq:preeomvacuumJT}) must hold locally. Consequently, the first law of complexity is equivalent to both the linearized gravitational and dilaton equations of motion holding in the bulk, i.e, $\delta E_g^{\mu \nu}=0$ and $\delta E_{\Phi} =0$. 

\subsubsection{Semi-classical corrections}

A novel feature of JT gravity is that it has fully analytic solutions even when 1-loop quantum effects are incorporated. This allows for a complete study of backreaction. These semi-classical effects are entirely captured by a non-local Polyakov action \cite{Polyakov:1981rd} and its associated Gibbons-Hawking-York boundary term 
\beq I_\chi = I^{\text{Poly}}_{\chi}+I_{\chi}^{\text{GHY}}=-\frac{c}{24\pi}\int_{\tilde{\mathcal{M}}} d^{2}x\sqrt{-g}\left[\chi R+(\nabla\chi)^{2}\right]-\frac{c}{12\pi}\int_{\mathcal{B}}dt\sqrt{-\gamma}\chi \mathcal{K}\;.\label{eq:semiclassactionconts}\eeq
Here $\chi$ is a local auxiliary field introduced such that the non-local Polyakov contribution appears local.\footnote{Substituting the formal solution $\chi =\frac{1}{2}\int d^{2}y\sqrt{-g(y)}G(x,y)R(y)$, where $G(x,y)$ is the Green's function for the D'Alembertian  $\Box$ operator, into the local form of the Polyakov action~(\ref{eq:semiclassactionconts}) one recovers the non-local  form,  $I_{\text{Poly}}=-\frac{c}{96 \pi}\int d^{2}x\sqrt{-g(x)}\int d^{2}y\sqrt{-g(y)}R(x)G(x,y)R(y)\;.$}
The Polyakov term is motivated by the two-dimensional conformal anomaly associated with a classical CFT of central charge $c$
\beq g^{\mu\nu}\langle T^{\chi}_{\mu\nu}\rangle=\frac{c}{24\pi}R\;.\eeq
The $\chi$ field models $c$ massless scalar fields (and is thus a CFT of central charge $c$), which can be taken to represent Hawking radiation of an evaporating black hole.

The equation of motion for $\chi$ is 
\beq 2\Box\chi=R\;.\label{eq:chieom}\eeq
The semi-classical action (\ref{eq:semiclassactionconts}) augments the classical gravitational equation  of motion (\ref{eq:gravEOMclassJT}) with backreaction fully characterized by $\langle T^{\chi}_{\mu\nu}\rangle\equiv-\frac{2}{\sqrt{-g}}\frac{\delta I_{\chi}}{\delta g^{\mu\nu}}$, such that the semi-classical gravitational equations  of motion are
\beq T_{\mu\nu}^{\phi}+\langle T^{\chi}_{\mu\nu}\rangle=0\;,\label{eq:semiclassgraveom}\eeq
with 
\beq \langle T^{\chi}_{\mu\nu}\rangle=\frac{c}{12\pi}\left[(g_{\mu\nu}\Box-\nabla_{\mu}\nabla_{\nu})\chi+(\nabla_{\mu}\chi)(\nabla_{\nu}\chi)-\frac{1}{2}g_{\mu\nu}(\nabla\chi)^{2}\right]\;.\label{eq:stresstenchi}\eeq
Here $\langle T^{\chi}_{\mu\nu}\rangle$ denotes the stress tensor expectation value with respect to some unspecified quantum state. The semi-classical approximation is only valid in the regime $\Phi_{0}/G\gg c\gg1$ \cite{Pedraza:2021cvx}.\footnote{The semi-classical approximation is understood as quantizing $c$ conformal fields, on a  classical background geometry. This is allowed since we ignore stringy-like ghost corrections to the dilaton or metric, provided $c\gg1$. However, for $c$ to be treated as a correction to the classical result, we impose $\Phi/G\gg c$.} Notice since $\chi$ does not couple to the dilaton directly, the dilaton equations of motion (\ref{eq:dilaEOMJT}) are unchanged, such that the background remains $\text{AdS}_{2}$. To solve the gravitational equations for $\Phi$ and $\chi$ one must first specify the quantum state of matter, which can lead to dramatically different solutions. For example, in the Hartle-Hawking state, the dilaton is only shifted by an overall constant proportional to $c$ (cf. \cite{Pedraza:2021cvx}).

\subsection*{Bulk quantum corrections to the first law}

The goal of this section is to compute the $1/N$ corrections to complexity=volume explicitly in two-dimensional dilaton gravity, where the dual $1/N$ corrections correspond to semi-classical bulk quantum corrections. We will see that, for an appropriate York deformation,  the variation of the holographic complexity formula gains a correction
\begin{equation}
    \delta_{\lambda_{f}} \mathcal{C} = \frac{1}{G\ell}\delta V_{JT} + \frac{i}{2}\int_{\mathcal{M}_{-}}  \langle \delta T^{\chi}_{\mu \nu} \rangle \delta_Y g^{\mu \nu}\;,
\label{eq:qcfirstlawJT}\end{equation} 
with the factor of $i$ due to $\mathcal{M}_{-}$ being in Euclidean signature, and can be absorbed via a Wick rotation back to Lorentzian signature. As we will argue below, it is natural to interpret this correction as a variation of `bulk complexity', i.e., the complexity due to bulk quantum fields $\chi$, $\delta c_{\text{bulk}}=\frac{1}{2}\langle \delta T^{\chi}_{\mu \nu} \rangle \delta_Y g^{\mu \nu}$. 

Realizing the quantum-corrected first law (\ref{eq:qcfirstlawJT}) requires a number of assumptions. First, we assume the boundary first law of complexity $\delta_{\lambda_{f}}\mathcal{C}=\Omega_{\text{bdry}}(\delta_{Y},\delta)$ exactly holds in the CFT, including $1/N$ quantum corrections. 
This assumption is motivated by the conjecture for the first law presented in \cite{Belin:2018bpg}, and studies of the first law of (Nielsen) complexity \cite{Bernamonti:2019zyy}. It also parallels the first law of entanglement entropy, which was used to show the universality of gravitation from entanglement even when $1/N$ corrections are included \cite{Swingle:2014uza,Agon:2021tia}.

Second, we are assuming matter perturbations $\delta\chi$ about vacuum $\text{AdS}_{2}$ (where $\chi=0$). Let us explore the consequences of this at the level of the symplectic structure. The bulk symplectic form of the complete semi-classical theory includes contributions from the Polyakov term, which add linearly to the classical symplectic form (\ref{eq:bulksympformclassJT}) (the same modification occurs when deriving the thermodynamic first law of quantum extremal surfaces \cite{Pedraza:2021cvx,Svesko:2022txo})
\beq \Omega_{\chi}(\phi,\delta_{1}\phi,\delta_{2}\phi)=\int_{\Sigma_{t}}(\delta_{1}\pi^{ab}_{\chi}\delta_{2}h_{ab}-\delta_{2}\pi^{ab}_{\chi}\delta_{1}h_{ab})+\int_{\Sigma_{t}}(\delta_{1}\pi_{\chi}\delta_{2}\chi-\delta_{2}\pi_{\chi}\delta_{1}\chi)\;,\label{eq:bulksympSJT}\eeq
where 
\beq \pi_{\chi}^{ab}\equiv\frac{\delta L_{\text{ADM}}^{\text{Poly}}}{\delta \dot{h}_{ab}}=\frac{c}{24\pi}\frac{\sqrt{h}}{N}(\dot{\chi}-N^{c}\partial_{c}\chi)h^{ab},\quad \pi_{\chi}\equiv\frac{\delta L_{\text{ADM}}^{\text{Poly}}}{\delta\dot{\chi}}=2\pi_{\chi}^{ab}h_{ab}+\frac{c}{12\pi}\sqrt{h}K,\eeq
are the momenta conjugate to $h_{ab}$ and $\chi$ in the Polyakov action alone (cf. (\ref{eq:conjmomgenapp})).\footnote{Since the field $\chi$ is non-minimally coupled to the two-dimensional Ricci scalar in the same way as the dilaton (and is not directly coupled to $\Phi$), the total momenta conjugate to $h_{ab}$ is given by the linear combination of the classical contribution (\ref{eq: dotphi}) and $\pi_{\chi}^{ab}$.} For deformations about vacuum AdS, the new York deformation (\ref{eq:nuNYtransJT}) acts as a diffeomorphism, such that $\delta_{Y}$ is extended so $\delta_{Y}\chi=0$, and, consequently, $\delta_{Y}\pi^{ab}_{\chi}=0$. Then, the action of $\delta_{Y}$ on the total bulk symplectic form for semi-classical JT gravity is
\beq \Omega_{\text{SJT}}(\phi,\delta_{Y}\phi,\delta\phi)=\alpha\delta V_{\text{JT}}+\int_{\Sigma}\delta_{Y}\pi_{\chi}\delta\chi\;,\label{eq:OmSJT2}\eeq
where $\delta_{Y}\pi_{\chi}=-\alpha c\sqrt{h}/12\pi$ via the deformation of configuration space variables (\ref{eq: configvariables}). In this process, we have uncovered a quantum-corrected volume, which we will discuss momentarily.

Thirdly, we assume, for simplicity, the matter perturbation $\delta \chi$ vanishes near the boundary. Together with the other assumptions, this results in the right-hand side of the first law (\ref{eq:qcfirstlawJT}). To see this, formally we express the bulk symplectic form as
\beq
\begin{split}
\Omega_{\text{SJT}}(\phi,\delta_{Y}\phi,\delta\phi)&=\frac{1}{G\ell}\delta V_{\text{JT}}+\int_{\Sigma}\omega_{\chi}(\phi,\delta_{Y}\phi,\delta\phi)\\
&=\frac{1}{G\ell}\delta V_{\text{JT}}-i\int_{\mathcal{M}_{-}}\hspace{-2mm}d\omega^{E}_{\chi}(\phi,\delta_{Y}\phi,\delta\phi)\;,
\end{split}
\eeq
where in the second line we Wick rotated to Euclidean signature, invoked Stokes' theorem, $\int_{\Sigma}\omega^{E}_{\chi}=\int_{\partial\mathcal{M}_{-}}\omega_{\chi}^{E}-\int_{\mathcal{M}_{-}}d\omega^{E}_{\chi}$, and that $\delta\chi$ vanishes at the (Euclidean) boundary. Using $\delta_{Y}$ acts like a diffeomorphism and the linearized equations of motions of the Polyakov action alone, we have
\beq d\omega_{\chi}^{E}(\phi,\delta_{Y}\phi,\delta\phi)=-\delta E^{\mu\nu}_{\chi}\delta_{Y}g_{\mu\nu}=-\frac{1}{2}\langle \delta T^{\mu\nu}_{\chi}\rangle\delta_{Y}g_{\mu\nu}\;.\eeq
Thus, 
\beq \Omega_{\text{SJT}}(\phi,\delta_{Y}\phi,\delta\phi)=\frac{1}{G\ell}\delta V_{\text{JT}}+\frac{i}{2}\int_{\mathcal{M}_{-}}\langle \delta T^{\mu\nu}_{\chi}\rangle\delta_{Y}g_{\mu\nu}\;.\label{eq:OmSJTv1}\eeq
Meanwhile, the boundary symplectic form is 
\beq
\begin{split}
 \Omega_{\text{bdry}}(\delta_{Y}\tilde{\lambda},\delta\tilde{\lambda})&=i\int_{\partial\mathcal{M}_{-}}\hspace{-2mm}(\omega^{E}_{\text{JT}}+\omega^{E}_{\chi})=i\int_{\Sigma}\omega_{\text{SJT}}^{E}+i\int_{\mathcal{M}_{-}}\hspace{-2mm}d\omega^{E}_{\text{SJT}}\;,
\end{split}
\label{eq:bdrysymSJT}\eeq
where the second equality follows from Stokes' theorem. When the background obeys the linearized equations of motion of the full semi-classical JT theory (together with $\delta_{Y}$ being a diffeomorphism), such that $d\omega_{\text{SJT}}^{E}=0$, and Wick rotating to Lorentzian signature, we recover $\Omega_{\text{bdry}}=\Omega_{\text{SJT}}$. Combined with (\ref{eq:OmSJTv1}), we recover the quantum-corrected first law (\ref{eq:qcfirstlawJT}).

Before describing how the semi-classical equations of motion emerge from the corrected holographic first law, let us briefly return to the semi-classical bulk symplectic form (\ref{eq:OmSJT2}). We are motivated to write a quantum-corrected volume $V_{\text{SJT}}$ where
\beq \delta V_{\text{SJT}}=\delta V_{\text{JT}}-\frac{c}{12\pi}\int_{\Sigma}dy\sqrt{h}\delta\chi\;.\eeq
This is almost the form of the volume functional one might have expected for semi-classical JT gravity. That is, recall the volume functional for classical JT (\ref{eq: jtvolume}) was motivated by the fact one expects complexity to scale as the entropy of the system. When quantum corrections are included, the (classical) gravitational entropy is modified by quantum corrections, such that the complexity scales as `generalized entropy', $S_{\text{gen}}=S_{\text{class}}+S_{\text{VN}}$, where $S_{\text{vN}}$ is von Neumann entropy due to bulk quantum fields. It happens that, in the case of semi-classical two-dimensional dilaton theories of gravity, on-shell the field $\chi$ is proportional to the von Neumann entropy of a two-dimensional CFT restricted to an interval, and the semi-classical Wald entropy is precisely equal to the generalized entropy \cite{Pedraza:2021cvx,Svesko:2022txo}
\beq S_{\text{Wald}}^{\text{SJT}}=\frac{1}{4G_{2}}(\Phi_{0}+\Phi)-\frac{c}{6}\chi=S_{\text{gen}}\;.\eeq
This suggests, at least in two dimensions, the volume functional (\ref{eq: jtvolume}) is modified to be
\beq V'_{\text{SJT}}=\int_{\Sigma}\sqrt{h}[(\Phi+\Phi_{0})+\gamma\chi]\;,\eeq
where $\gamma$ is some constant proportional to $c$ such that $(\Phi+\Phi_{0})\gg\gamma\chi$. We can now ask what is the new York deformation such that 
\beq \Omega_{\text{SJT}}(\phi,\delta_{Y}\phi,\delta\phi)=\alpha\delta V'_{\text{SJT}}=\alpha\int_{\Sigma_{t}}dy[(\delta\sqrt{h})(\Phi+\Phi_{0}+\gamma\chi)+\sqrt{h}(\delta\Phi+\gamma\delta\chi)]\;.\label{eq:OmSJTv3}\eeq
Naively, it is natural to simply extend the classical deformation (\ref{eq:nuNYtransJT}) to also require
 \beq \delta_{Y}\pi^{ab}_{\chi}=\frac{\alpha\gamma}{2}\sqrt{h}h^{ab}\chi\;,\quad \delta_{Y}\pi_{\chi}=\alpha\gamma\sqrt{h}\;,\quad \delta_{Y}\chi=0\;.\eeq
However, the deformations $\delta_{Y}\pi_{\chi}^{ab}$ and $\delta_{Y}\chi$ are incompatible, given the phase space variables (\ref{eq:bulksympSJT}). Consequently, one must modify the deformation used for the classical JT gravity. This would be interesting when one considers perturbations about the \emph{backreacted} background, i.e., the initial background includes $\chi$. One would then look for a different new York deformation such that (\ref{eq:OmSJTv3}) holds and the Hamiltonian and momentum constraints of semi-classical JT gravity are preserved. In our case, we are only concerned with perturbations to the vacuum background, such that the York deformation need only preserve the Hamiltonian and momentum constraints of classical JT gravity. Since the semi-classical theory is exactly solvable, it would be interesting to study perturbations to the backreacted background.

\subsection*{Emergence of linearized semi-classical JT gravity}

Having established the quantum-corrected holographic first law of complexity (\ref{eq:qcfirstlawJT}), we can once again reverse engineer to derive the linearized equations of motion for semi-classical JT gravity. The same logical steps leading to the integral identity \eqref{eq: pushingVolumeToBoundary} are used. More precisely, by Stokes' theorem we have
\beq 
\begin{split}
i\int_{\mathcal{M}_{-}}d\omega_{\text{SJT}}^{E}&=i\left(\int_{\partial\mathcal{M}_{-}}\omega_{\text{SJT}}^{E}(\phi,\delta_{Y}\phi,\delta\phi)-\int_{\Sigma}\omega^{E}_{\text{SJT}}(\phi,\delta_{Y}\phi,\delta\phi)\right)\\
&=\Omega_{\text{bdry}}(\delta_{Y}\tilde{\lambda},\delta\tilde{\lambda})-\left(\frac{1}{G \ell}\delta V_{\text{JT}}+\frac{i}{2}\int_{\mathcal{M}_{-}}\langle \delta T^{\mu\nu}_{\chi}\rangle\delta_{Y}g_{\mu\nu}\right)\;,
\end{split}
\eeq
where in the second line we invoked the bulk and boundary symplectic forms (\ref{eq:OmSJTv1}) and (\ref{eq:bdrysymSJT}). With $\delta_{\lambda_{f}}\mathcal{C}=\Omega_{\text{bdry}}(\delta_{Y}\tilde{\lambda},\delta\tilde{\lambda})$, imposing the quantum-corrected holographic first law (\ref{eq:qcfirstlawJT}), ultimately leads to the linearized equations of motion for the metric and dilaton fields for bulk matter perturbations about the classical background. In reference to the triangle depicted in Figure \ref{fig:trinity}, we have shown, the boundary first law together with its holographic realization, including quantum corrections, gives rise to the semi-classical equations of motion. Motivated by this, let us formally consider the effects of bulk quantum corrections in higher-dimensional theories of gravity.

\subsection{A proposal for bulk complexity in higher dimensions} \label{sec:quantumcorrectionsprop}

Above we showed semi-classical equations of motion of JT gravity follow from the holographic first law of CV complexity, including a suitable generalization including a contribution over the bulk $\mathcal{M}_{-}$, in addition to the volume. The bulk contribution arises solely due to semi-classical backreaction effects. It is natural to interpret this additional term as a `bulk complexity' characterizing the complexity of bulk quantum fields. We will say more about this interpretation momentarily, but for now, the point is semi-classical gravitational equations of motion arise from assuming the quantum-corrected first law of complexity. In other words, there is a type of universality of gravity from complexity. This is a complementary viewpoint from holographic entanglement: when $1/N$ corrections to the dual CFT are included,  the bulk side of the RT formula is supplemented by entanglement entropy due to bulk quantum fields \cite{Faulkner:2013ana}, and, moreover, the quantum-corrected first law of entanglement is equivalent to imposing semi-classical equations of motion in the bulk \cite{Swingle:2014uza,Agon:2021tia}. Note that, assuming the universality of gravity, one could have predicted the form of the bulk quantum correction to the classical area law, starting with semi-classical equations of motion as an input.

To expound on this point, let us recall the derivation of the semi-classical Einstein equations from the first law of (holographic) entanglement. In general QFTs, the von Neumann entropy $S_{A}$ of a quantum state $\rho_{A}$ restricted to a domain $A$ obeys a first law
\beq \delta S_{A}=\text{tr}(\delta\rho_{A}H_{A})=\delta\langle H_{A}\rangle\;,\eeq
under small perturbations around a reference state, $\rho_{A}=\rho_{A}^{(0)}+\delta\rho_{A}$.  Here $H_{A}$ refers to the modular Hamiltonian, formally defined through expressing the state in Gibbs form, $\rho_{A}=e^{-H_{A}}/\text{tr}e^{-H_{A}}$. In the event $\rho_{A}$ is an actual thermal state, the first law of entanglement represents a quantum version of the first law of thermodynamics, $\delta S_{A}=\delta E_{A}$, where $E_{A}=\langle H_{A}\rangle$ is some energy corresponding to subsystem $A$. Such examples include the vacuum state of a QFT restricted to a Rindler wedge, or a ball, where $H_{A}$ acts as the generator of time-translations of Rindler observers and is expressed in terms of an integral of the time-time component of the QFT stress-tensor over $A$. For holographic CFTs, where the Ryu-Takayanagi prescription applies, there exists a $(d-1)$-form $\tilde{\chi}$ in bulk $\text{AdS}_{d}$ such that 
\beq \int_{A}\tilde{\chi}=\delta E_{A}^{\text{grav}}\;,\quad \int_{\gamma_{A}}\tilde{\chi}=\delta S^{\text{grav}}_{A}\;,\label{eq:chiformdef}\eeq
and $d\tilde{\chi}=-2\xi^{\mu}\delta E^{g}_{\mu\nu}\epsilon^{\nu}$. Here $\delta E^{\text{grav}}_{A}$ and $\delta S_{A}^{\text{grav}}$ are gravitational counterparts of $\delta E_{A}$ and $\delta S_{A}$, $\gamma_{A}$ is the RT surface homologous to $A$, $\xi$ is a time-like conformal Killing vector, $\epsilon^{\nu}$ is a volume form on a codimension-1 bulk region $\Sigma_{A}$, and $\delta E^{g}_{\mu\nu}$ is the linearized gravitational equations of motion for small bulk perturbations. Clearly, the form $\tilde{\chi}$ is closed for on-shell perturbations, such that the holographic first law of entanglement is satisfied $\delta S^{\text{grav}}_{A}=\delta E^{\text{grav}}_{A}$. Alternatively, assuming the first law of holographic entanglement, one finds $d\tilde{\chi}$ integrates to zero such that $\delta E^{g}_{\mu\nu}=0$ locally in the bulk \cite{Faulkner:2013ica,Agon:2020mvu}. 

Now include the effects of semi-classical bulk quantum corrections. Backreaction due to bulk quantum fields induces an order $O(G^{0})$ change to the metric such that 
\beq \delta S_{A}^{\text{grav}}=\delta S_{A}^{\text{grav,class}}+\delta S_{\text{bulk}}(\Sigma_{A})\;,\label{eq:bulkentvar}\eeq
where $\delta S_{A}^{\text{grav,class}}$ represents the usual geometric contribution to classical gravitational entropy, e.g., the variation of the area of the bulk entangling surface in the case of Einstein gravity. Moreover, for bulk regions $\Sigma_{A}$ with a local modular Hamiltonian, the variation of bulk entropy obeys $\delta S_{\text{bulk}}(\Sigma_{A})=\int_{\Sigma_{A}}\xi^{\mu}\langle \delta T_{\mu\nu}^{\text{bulk}}\rangle\epsilon^{\nu}$.
Consequently, 
\beq \delta S^{\text{grav}}_{A}=\int_{\gamma_{A}}\tilde{\chi}+\int_{\Sigma_{A}}\xi^{\mu}\langle \delta T_{\mu\nu}^{\text{bulk}}\rangle\epsilon^{\nu}\;.\eeq
Meanwhile, for bulk perturbations that decay sufficiently fast near the boundary, $\delta E^{\text{grav}}_{A}$ is unaltered, such that the variational expression in (\ref{eq:chiformdef}) is satisfied. Then, by Stokes' theorem,
\beq \delta S_{A}^{\text{grav}}-\delta E^{\text{grav}}_{A}=\int_{\Sigma_{A}}[d\tilde{\chi}+\xi^{\mu}\epsilon^{\nu}\langle \delta T_{\mu\nu}^{\text{bulk}}\rangle]=0\;.\eeq
Thus, when the first law of entanglement holds  one finds that the bulk satisfies the linearized semi-classical gravitational equations of motion. Suppose, alternatively, we did not know precisely how to express the variation of the bulk entanglement entropy, but we knew bulk entropy variation (\ref{eq:bulkentvar}). Using the universality of spacetime entanglement as a guiding principle, such that the holographic first law of entanglement implies semi-classical bulk equations of motion, we would conclude $\delta S_{\text{bulk}}(\Sigma_{A})=\int_{\Sigma_{A}}\delta s^{\text{bulk}}(\Sigma_{A})=\int_{\Sigma_{A}}\xi^{\mu}\langle \delta T_{\mu\nu}^{\text{bulk}}\rangle\epsilon^{\nu}$.

Motivated by this observation, and our explicit computation of the holographic first law of complexity in semi-classical JT gravity, we now speculate on the form the bulk complexity must take for arbitrary higher-dimensional theories of gravity assuming the universality of gravity from complexity. This is by no means a proof of the quantum-corrected formula for holographic complexity, but rather a formal determination, assuming the principle of spacetime complexity extends to subleading $1/N$ corrections in the field theory. 

To this end, we start by assuming the quantum-corrected first law of holographic (CV) complexity takes the following general form (in Euclidean signature)
\begin{equation}\label{eq: higherCorrections}
\delta \mathcal{C}= \frac{1}{G \ell} \delta W_{\text{gen}} +i \int_{\mathcal{M}_{-}} \delta c_{\text{bulk}}.
\end{equation}
Here $\delta c_{\text{bulk}}$ represents a bulk contribution to complexity due to bulk quantum fields. Such bulk quantum corrections have been proposed before in the context of holographic braneworlds \cite{Hernandez:2020nem}, such that generally,
\beq \mathcal{C}=\frac{1}{G\ell}\,\underset{\Sigma\sim A}{\text{max}}\left[W_{\mathrm{gen}}(\Sigma)+W_{K}(\Sigma)+C_{\text{bulk}}\right]\;.\eeq
Our goal is to find the form of $\delta c_{\text{bulk}}$ such that semi-classical bulk equations of motion are imposed in the bulk, at least for perturbative states around the AdS vacuum. For simplicity, we assume the bulk matter fields are minimally coupled. 

We start by assuming the boundary first law
\beq \delta_{\lambda_{f}}\mathcal{C}=\Omega_{\text{bdry}}(\delta_{Y}\tilde{\lambda},\delta\tilde{\lambda})\;.\eeq
Holographically we have,
\begin{equation}
   \Omega_{\text{bdry}}(\delta_{Y}\tilde{\lambda},\delta\tilde{\lambda}) =  \frac{1}{G\ell} \delta W_{\text{gen}}+ i \int_{\mathcal{M}_{-}} \delta c_{\text{bulk}},
\end{equation}
such that we have decomposed $\Omega_{\text{bulk}}=\Omega_{\text{bulk}}^{\text{grav}}+\delta C_{\text{bulk}}$, with $\Omega_{\text{bulk}}^{\text{grav}}$ being the purely gravitational contribution to the bulk symplectic form, the analog of $\delta S_{A}^{\text{grav,class}}$.
Now reverse engineer the steps which led to the integral identity \eqref{eq: pushingVolumeToBoundary}. We find
\begin{equation}\label{eq: bulkcondition}
    \int_{\mathcal{M}_{-}} (d \omega_{\text{bulk}}^{E}(\delta_Y \phi, \delta \phi) - \delta c_{\text{bulk}}) = 0,
\end{equation}
for bulk dynamical fields $\phi=\{g,\chi\}$. Recall that, when no matter is present, about the AdS vacuum the new York deformation $\delta_{Y}$ acts as a diffeomorphism. Assuming matter perturbations $\delta\chi$ which vanish sufficiently fast near the boundary and extended the action of $\delta_{Y}$ on bulk fields $\chi$ such that $\delta_{Y}\chi=\mathcal{L}_{Y}\chi$ (as in semi-classical JT gravity), then
\begin{equation}\label{eq: omegaderivation}
    d \omega_{\text{bulk}}^{E} (\delta_Y \phi, \delta \phi ) = -\delta_Y g_{\mu \nu} \delta E_{g}^{\mu \nu},
\end{equation}
where as usual $\delta E_{g}^{\mu\nu}$ denotes the gravitational component of the linearized bulk equations of motion. Consequently, in order for the semi-classical equations of motion be satisfied, we arrive at our proposal for the subleading quantum corrections to holographic complexity,
\begin{equation}
    \delta c_{\text{bulk}} =  \frac{1}{2} \delta_Y g_{\mu \nu} \langle \delta T_{\text{bulk}}^{\mu \nu} \rangle,
\end{equation}
where $\langle \delta T_{\text{bulk}}^{\mu\nu}\rangle$ represents the variation of the stress-energy tensor of bulk quantum fields. 

Let us take stock in the implicit physical assumptions made to arrive at this result. First, we assumed that there are two contributions to the total bulk symplectic form $\Omega_{\text{bulk}}(\delta_{Y},\delta)$: a purely gravitational contribution, $\Omega_{\text{bulk}}^{\text{grav}}$, and another due to complexity of bulk fields. Notably, the gravitational component of the bulk symplectic form is of order $O(G^{0})$. To see this, we adapt the argument of \cite{Swingle:2014uza,Agon:2021tia} for our context. We are interested in (coherent) CFT states describing a bulk semi-classical spacetime with a bulk quantum field in state $|\psi\rangle_{\text{bulk}}$. For perturbative excited states, i.e., perturbations about vacuum AdS, the expectation value $\langle\psi|T_{\mu\nu}^{\text{bulk}}|\psi\rangle$ is generally non-vanishing. Correspondingly, this stress-tensor will source the semi-classical equations of motion, inducing a metric perturbation due to backreaction, $g_{\mu\nu}=g^{(0)}_{\mu\nu}+\delta g_{\mu\nu}$, with $\delta g_{\mu\nu}\sim O(G)$. Hence, the volume form $\epsilon_{\Sigma}$ on the (maximal) bulk slice $\Sigma$ will receive a correction of the same order (while the surface $\Sigma$ itself is not expected to change at linear order). Thus, $\frac{1}{G\ell}\delta W_{\text{gen}}\sim O(G^{0})$. Clearly, when backreaction is not accounted for, i.e., for \emph{classical} perturbative states, the volume only receives corrections of order $O(G^{-1})$. This term dominates over the bulk complexity contribution, which must thus enter at order $O(G^{0})$. In other words, for semi-classical states, both $\Omega_{\text{bulk}}^{\text{grav}}$ and the bulk complexity, $\delta C_{\text{bulk}}$ enter at the same order in a $G$ expansion. Second, we considered bulk perturbations which decay sufficiently fast near the conformal boundary, as done in the context of bulk quantum corrections to holographic entanglement \cite{Swingle:2014uza,Agon:2021tia}. 

At this stage we are unable to say more about the precise nature of bulk complexity. Loosely speaking, in the language we have employed here, it is natural to think of bulk complexity in terms of bulk state preparation. That is, there exist a set of bulk sources which are used to specify a bulk quantum state via a Hartle-Hawking prescription. It would be interesting to make this more precise using either a geometric approach as we have done here, or by developing a notion of bulk path integral optimization \cite{Chandra:2022pgl}. 

\section{Discussion} \label{sec:disc}

Assuming a particular notion of boundary complexity, with suitable generalizations of the CV dictionary, we have shown linearized gravitational equations of motion emerge from varying complexity. In particular, we derived gravitational dynamics for any higher-order theory of pure gravity by replacing the standard volume of the maximal slice with the generalized volume. Moreover, when bulk quantum corrections are included, the holographic first law acquires a correction to the geometric contribution, which we refer to as bulk complexity, an analog of bulk entanglement entropy appearing in the generalization of the Ryu-Takayanagi formula. While we determined the form of bulk complexity via explicit computation in the context of semi-classical JT gravity, we postulate such a term should appear in more general theories. In fact, such a term is necessary if one hopes to recover semi-classical equations of motion from varying complexity. Collectively, our work deepens the connection between gravity and complexity, beyond only probing the interior of black holes at late times. Rather, gravity appears to have a quantum computational origin. 

We emphasize we have derived the semi-classical equations of motion, despite AdS/CFT being an example of a complete quantum theory of gravity. This is because our derivation was strictly concerned with coherent states of holographic CFTs. Notably, coherent states of a quantum system have dynamics which closely resemble a classical state. Holographically, these coherent states map to spacetimes with a `good' semi-classical description, i.e., a single classical geometric background with quantum fields on top. This implies, as in the case of the Ryu-Takayanagi prescription for holographic entanglement, that the standard complexity=volume conjecture is only valid within a semi-classical approximation. That is, volume complexity is seemingly ill-defined with respect to quantum states whose bulk dual is described as a superposition of geometries.\footnote{In the context of JT gravity, however, a path integral definition of the length of an Einstein-Rosen bridge exists, including a sum over metrics (codified as boundary `wiggles') and non-perturbative corrections \cite{Iliesiu:2021ari}. Its early and late-time asymptotics behaves as expected, suggesting a quantum gravitational notion of CV.}

Finally, having derived gravitational equations of motion including bulk matter contributions, our work suggests the universality of optimized computation is intimately linked to the universality of gravity. That is, by turning on sources dual to light fields (in a large-$c$ CFT with a holographic dual), the notion of complexity we work with changes universally, dictated by the fact all matter sources gravity in a universal manner. Turning this observation around, the universality of computation -- all paths in the space of sources contribute to system dynamics -- leads, via holography, to the universality of gravitational interaction. 

\vspace{2mm}

There are multiple interesting future directions to take our work, some of which address caveats we encountered above. Let us describe these future directions in some detail.

\vspace{2mm}

\noindent \textbf{Beyond complexity=volume.} In this article we established an equivalence between the first law of boundary complexity, its holographic description, and linearized bulk equations of motion. We showed that given the first two of these inputs, we were able to reconstruct the third. Alternately, one could assume the equations of motion and construct one of the other inputs. For example, in our approach, we committed ourselves to a specific form of boundary first law, motivated by the form of complexity in \eqref{eq:compftdef} where the associated cost function is taken to be the kinetic energy in the space of sources of coherent states. An \emph{ad hoc} motivation for this choice was such that the change of bulk volume or generalized volume follows from linear variations in sources. It would be natural to consider other types of cost functions in the space of sources. In so doing, it is not expected linear variations of sources would lead to a change in volume, but would be related to some other bulk quantity. This connection between the inherent freedom in choosing cost functions and the holographic dual of complexity has been explored in \cite{Belin:2021bga,Belin:2022xmt}, leading to the proposal `complexity=anything'.  In this approach, one considers the (infinite) class of functionals such that the late-time behavior and switchback effect are satisfied, without imposing any other criterion. Our work here may provide another criterion, namely, the emergence of bulk dynamics, as a means to further restrict the class of functionals. 

Further, a motivation for the `complexity=anything' proposal is that, generally, complexity is innately ambiguous due to a choice in cost function. It would be interesting to consider other notions of boundary complexity by constructing other types of cost functions and see how these notions relate to different bulk functionals, by requiring consistency of bulk dynamics. To this end, perhaps path integral constructions could be used \cite{Chandra:2021kdv,Chandra:2022pgl}. Moreover, it would be worth exploring the connection between the geometry of the space of sources preparing coherent states and the choice of cost function. One possible route is to rephrase our language using the machinery of generalized free fields, where the role of the boundary symplectic form may correspond to a symplectic bilinear which itself is ambiguous (for a recent discussion on these matters, see \cite{Furuya:2023fei}).\footnote{We thank Nima Lashkari for explaining this point to us.}

\vspace{2mm}

\noindent \textbf{Non-linear corrections.} Here we derived the linearized equations of motion of arbitrary higher-order gravity theories, specifically for perturbations about empty AdS. As in the case of deriving gravitational equations of motion from entanglement \cite{Faulkner:2017tkh}, it is expected perturbations about excited CFT states carry information about non-linear corrections, corresponding to second-order contributions to sources $\lambda$ in the Euclidean path integral definition of an excited CFT state. Thus, second-order variations to the complexity should capture non-linear corrections to gravitational equations of motion. It would be interesting to determine the form of these corrections for the class of coherent states studied in this paper.

\vspace{2mm}

\noindent \textbf{State preparation and complexity on holographic braneworlds.} To explore the effects of $1/N$ quantum corrections to holographic complexity, we focused on an example of two-dimensional dilaton gravity where semi-classical quantum backreaction effects can be exactly incorporated. The only other framework where it is known how to exactly account for quantum backreaction is braneworld holography \cite{deHaro:2000wj}, which has been successful in uncovering fully backreacted `quantum' black holes \cite{Emparan:2020znc,Emparan:2022ijy,Panella:2023lsi}. In fact, the generalized complexity formula, including bulk quantum corrections, was proposed based on a braneworld construction \cite{Hernandez:2020nem}, and the effects of bulk quantum corrections to the complexity of the quantum BTZ black hole were analyzed in \cite{Emparan:2021hyr}. It would be interesting to consider holographic state preparation in the context of such `doubly holographic' braneworld models. A first step in this direction would require an understanding of how to prepare coherent boundary CFT (BCFT) states via a Euclidean path integral with sources. Then, via the layered holography, this should produce a dual description of state preparation of bulk states including a brane. Sources residing on the conformal boundary of bulk (Euclidean) AdS would then be identified as sources on the brane. From the brane perspective, such sources would then be responsible for a type of bulk complexity, and, in principle, be known to all orders in $1/N$.  

\vspace{2mm}

\noindent \textbf{Subregion complexity and symplectic structure.} Here we considered the complexity of dual states defined on a global timeslice. This version of complexity=volume has been extended to `subregion complexity', analyzing the complexity of a dual CFT state reduced to a boundary subregion. Namely, the complexity of a quantum state defined on a boundary subregion is given by the volume of a maximal codimension-1 bulk surface between the subregion and associated bulk RT entangling surface \cite{Alishahiha:2015rta,Carmi:2016wjl} (in fact, this was the set-up considered in the holographic braneworld models \cite{Hernandez:2020nem,Bhattacharya:2021jrn}). It would be interesting to extend our analysis to incorporate proposals of subregion complexity and explore the relation between boundary and bulk symplectic structures in this context. A starting point is to frame subregion complexity in terms of holographic state preparation, i.e., how to prepare a target state reduced to a boundary subregion. An analogous problem, and motivation for subregion complexity, appears in entanglement wedge reconstruction, which exploits the duality between boundary and bulk modular flow. Recently it was shown the expectation value of the modular Berry curvature for a broad class of coherent CFT state deformations -- defined via Euclidean path integrals with sources -- is dual to a suitably defined bulk symplectic form of the entanglement wedge \cite{Czech:2023zmq}. It would be worth exploring the connection between \cite{Czech:2023zmq} and our work to develop our understanding of subregion complexity.

\vspace{2mm}

\noindent \textbf{Lorentzian flow interpretation.} Via a flow-based reformulation of CV complexity \cite{Pedraza:2021fgp,Pedraza:2021mkh}, a conceptual picture of spacetime complexity emerges. In this view, maximization of volume is replaced by minimization of Lorentzian `threads' -- divergenceless, timelike vector fields -- which begin on the boundary $\partial\mathcal{M}_{-}$ of the southern hemisphere, with each thread attached to a source $\lambda_{f}$ characterizing the reference state used to prepare a target state at $\Sigma$. In this language, complexity is equal to the minimum number of threads passing through $\Sigma$. When the CFT states are described in terms of tensor networks, an apt visualization follows: Lorentzian threads sew together tensor networks discretizing slices foliating the bulk spacetime. Moreover, this flow-based point of view is consistent with spacetime dynamics emerging from varying complexity. This is because, a natural choice for an optimal Lorentzian thread configuration characterizing perturbations about empty AdS is the symplectic current $\omega_{\text{bulk}}(\delta_{Y},\delta)$, where the divergenceless condition coincides with $d\omega_{\text{bulk}}=0$, thereby imposing the linearized field equations. This relation between flows and the closedness of the bulk symplectic current was established for CFTs dual to Einstein gravity, however, it would be interesting to generalize to the case of higher-order theories. 

A conceptual puzzle regarding Lorentzian flows is: do Lorentzian threads `commute'? This is important because entanglement buildup often requires non-commuting unitary gates in a tensor network model. The divergenceless condition of threads (now interpreted as `gatelines'), meanwhile, suggests individual gates commute. It would thus be worth providing a flow-based view of `quantum Lorentzian threads', accounting for bulk quantum corrections, where the divergenceless condition is expected to be relaxed, as in the case of holographic entanglement entropy \cite{Agon:2021tia,Rolph:2021hgz}.

\section*{Acknowledgments}

We are grateful to José Barbón, Elena Cáceres, César Gómez, Michal Heller, Juan Hernandez, Nima Lashkari, Bahman Najian, Ayan Patra, Andrew Rolph, Brian Swingle, Jeremy van der Heijden, and Claire Zukowski for useful discussions and comments on the manuscript. RC and JFP are supported by the `Atracci\'on de Talento' program grant 2020-T1/TIC-20495 and by the Spanish Research Agency via grants CEX2020-001007-S and PID2021-123017NB-I00, funded by MCIN/AEI/10.13039/501100011033 and by ERDF A way of making Europe. AS is supported by the Simons Foundation via the \emph{It from Qubit Collaboration} and by EPSRC. AS and ZWD acknowledge IFT-Madrid for hospitality while this work was being completed.

\appendix

\section{The new York deformation in vacuum AdS} \label{app:NYVaccumAdS}

\textbf{Wheeler-de Witt coordinates.} In empty bulk AdS, the new York transformation $\delta_{Y}$ behaves as a diffeomorphism \cite{Belin:2018bpg}. One way to see this is to express a constant mean curvature slicing (CMC) of $\text{AdS}_{d}$ using Wheeler-de Witt (WdW) coordinates:
\beq ds^{2}_{d}=g_{\mu\nu}(x)dx^{\mu}dx^{\nu}=-d\tau^{2}+h_{ab}(\tau,y)dy^{a}dy^{b}\;,\label{eq:wdwcoordapp}\eeq
with $h_{ab}(\tau,y)=\cos^{2}(\tau)\sigma_{ab}(y)$ and determinant $h(\tau,y)=\cos^{2(d-1)}(\tau) \sigma(y)$. This foliation has constant time $\tau$ slices $\Sigma_{\tau}$. The timelike unit normal $n_{\alpha}=\delta^{\tau}_{\alpha}$ and the extrinsic curvature pulled back to $\Sigma_{\tau}$ is $K_{ab}=\frac{1}{2}\partial_{\tau}(\cos^{2}\tau)\sigma_{ab}(y)$, such that $K=-(d-1)\tan\tau$. Clearly, the $\tau=0$ slice is the maximal volume slice with $K=0$. Moreover, 
\beq \pi_{V}=-\frac{2(d-2)}{(d-1)}K=2(d-2)\tan\tau\;,\quad \partial_{\pi_{V}}=\frac{1}{2(d-2)}\cos^{2}\tau\partial_{\tau}\;.\eeq
Thus, a translation in York time $\pi_{V}$ corresponds to a translation time $\tau$.

In WdW coordinates, the conformal metric, $\bar{h}_{ab}(\tau,y)=|h(y)|^{-1/(d-1)}h_{ab}(y)$, is $\tau$-independent such that $\partial_{\tau}\bar{h}_{ab}=0$. It is also straightforward to verify $\partial_{\tau}\bar{\pi}_{ab}=0$ for any $\tau$. Only on the maximal $\tau=0$ slice, however, does $\partial_{\tau}\sqrt{h(\tau,y)}=0$. Next, define the spacetime vector field 
\beq \xi\equiv 2\alpha(d-2)\partial_{\pi_{V}}=\alpha\cos^{2}\tau\partial_{\tau}\;,\label{eq:Yorkvecwdw}\eeq
for some real parameter $\alpha$. It follows $\mathcal{L}_{\xi}\pi_{V}=\xi^{\mu}\partial_{\mu}\pi_{V}=2\alpha(d-2)$, and $\mathcal{L}_{\xi}\bar{h}_{ab}=\mathcal{L}_{\xi}\bar{\pi}_{ab}=0$ for any time $\tau$. Meanwhile,
\beq \mathcal{L}_{\xi}h_{ab}=-2\alpha \cos(\tau)\sin(\tau)h_{ab}(\tau,y)\;,\quad \mathcal{L}_{\xi}K_{ab}=-\alpha\cos(2\tau)h_{ab}(\tau,y)\;.\eeq
Combined, at the initial data ($\tau=0$) surface, $\mathcal{L}_{\xi}\phi=\delta_{Y}\phi$ satisfying the York deformations (\ref{eq:Yorkvariations}) and (\ref{eq: yorkTransformation}), and the new York transformation behaves as a diffeomorphism in vacuum AdS. In fact, since $\delta_{\xi}\sqrt{h}=0$ is only enforced on the $\tau=0$ slice, the spacetime vector field generating new York deformation is $\xi_{\tau=0}\equiv Y=\alpha\partial_{\tau}$. Then,
\beq \mathcal{L}_{Y}h_{ab}=-2\alpha\tan(\tau)h_{ab}(\tau,y)\;,\quad \mathcal{L}_{Y}K_{ab}=-\alpha(1-\tan^{2}(\tau))h_{ab}(\tau,y)\;,\eeq
where $\mathcal{L}_{Y}h_{ab}=0$ and $\mathcal{L}_{Y}K_{ab}=-\alpha h_{ab}$ at $\tau=0$.

\vspace{2mm}
 
\noindent \textbf{Poincar\'e coordinates.} It also useful to know the form of $\delta_{Y}$ in Poincar\'e patch coordinates. To this end, WdW coordinates (\ref{eq:wdwcoordapp}), with
\beq ds^{2}_{d}=-d\tau^{2}+\cos^{2}(\tau)\frac{du^{2}+d\tilde{x}^{2}}{u^{2}}\;,\eeq
may be transformed into the Poincar\'e patch
\beq ds^{2}_{d}=\frac{1}{z^{2}}(-dt^{2}+dz^{2}+d\tilde{x}^{2})\;,\eeq
via
\beq \tau=\arcsin(t/z)\;,\quad u=\sqrt{z^{2}-t^{2}}\;.\eeq
The inverse coordinate transformation is
\beq t=u\tan(\tau)\;,\quad z=u\sec(\tau)\;.\eeq
The $\tau=0$ slice coincides with the $t=0$ zero hypersurface. Meanwhile, the AdS conformal boundary at $z=0$ corresponds to $\tau=\pm i\infty$ in WdW coordinates. Using the inverse coordinate transformation, the spacetime vector field $Y=\alpha\partial_{\tau}$ transforms as
\beq Y=\frac{\alpha z^{2}}{\sqrt{z^{2}-t^{2}}}\partial_{t}+\frac{\alpha zt}{\sqrt{z^{2}-t^{2}}}\partial_{z}\;.\label{eq:YspacevecfieldGR}\eeq

\vspace{2mm}
 
\noindent \textbf{Euclidean signature.} In Euclidean signature, where $\tau\to-i\tau_{E}$ with $\tau_{E}\in\mathbb{R}$, the York deformation vector $Y\to i\alpha \partial_{\tau_{E}}\equiv iY_{E}$. Then, 
\beq \mathcal{L}_{Y}h_{ab}=i\alpha\partial_{\tau_{E}}h_{ab}=2i\alpha \tanh(\tau_{E})h_{ab}(\tau_{E},y)\;.\eeq
Moreover, in Poincar\'e coordinates, where, $t\to-it_{E}$ with $t_{E}=u\tanh(\tau_{E})$, we find
\beq Y\to iY_{E}\;,\quad Y_{E}=\frac{\alpha z^{2}}{\sqrt{z^{2}+t_{E}^{2}}}\partial_{t_{E}}-\frac{\alpha zt_{E}}{\sqrt{z^{2}+t_{E}^{2}}}\partial_{z}\;,\label{eq:NYtransEapp}\eeq
and
\beq 
\begin{split}
&\mathcal{L}_{Y}g_{t_{E}t_{E}}=\frac{2i\alpha t_{E}^{3}}{z^{2}(z^{2}+t^{2}_{E})^{3/2}}\;,\quad \mathcal{L}_{Y}g_{zz}=\frac{2i\alpha t_{E}}{(z^{2}+t_{E}^{2})^{3/2}}\;,\\
&\mathcal{L}_{Y}g_{z\tau}=\frac{2i\alpha t_{E}^{2}}{z(z^{2}+t_{E}^{2})^{3/2}}\;,\quad \mathcal{L}_{Y}g_{ij}=\frac{2i\alpha t_{E}}{z^{2}\sqrt{z^{2}+t_{E}^{2}}}\delta_{ij}\;.
\end{split}
\label{eq:NYtransEbulkP}\eeq
Near the boundary $z\to0$, the new York transformation (\ref{eq:NYtransEapp}) goes like $Y|_{z\to0}=-i\alpha z\text{sign}(t_{E})\partial_{z}$, such that $\mathcal{L}_{Y}\gamma_{\mu\nu}=2i\alpha \text{sign}(t_{E})\gamma_{\mu\nu}$, where $\gamma_{\mu\nu}$ is the boundary metric at $z=0$.\footnote{The action of $\delta_{Y}$ on the conformal boundary metric may be read off from taking the $z\to0$ limit of the $t_{E}t_{E}$ and $ij$ components of (\ref{eq:NYtransEbulkP}).} In this way, the new York transformation behaves as a Weyl rescaling of the boundary metric with Weyl factor $(1-2i\alpha\text{sign}(t_{E}))$.

\section{Bulk symplectic form in higher-order gravity: details}\label{app:generalized volume}

Here we provide some additional computational details leading to $\Omega_{\text{bulk}}(g,\delta_{Y}g,\delta g) \sim\delta W_{\text{gen}}$. We work with arbitrary diffeomorphism covariant theories of gravity with Lagrangian $d$-form\footnote{Here our abstract index notation uses Latin letters for the full spacetime index instead of Greek indices.}
\beq
    L=L(g_{ab},R_{bcde},\nabla_{a_1} R_{bcde},...,\nabla_{(a_1}...\nabla_{a_m)}R_{bcde}).
\eeq
For such a theory, the symplectic potential $\theta(g,\delta g)$ and symplectic current $\omega\equiv\delta_{1}\theta(g,\delta_{2}g)-\delta_{2}\theta(g,\delta_{1}g)$ $(d-1)$-forms may be cast as (see \cite{Iyer:1994ys})
\beq \theta(g,\delta g)=2P^{bcd}\nabla_{d}\delta g_{bc}+S^{ab}\delta g_{ab}+\sum_{i=1}^{m-1}T_{i}^{abcda_{1}...a_{i}}\delta\nabla_{(a_{1}}...\nabla_{a_{i})}R_{abcd}\;,\eeq
\begin{equation}
\begin{split}
        \omega_{\text{bulk}}(g,\delta_1 g,\delta_2 g)=&2 \delta_1 P^{bcd}\nabla_d\delta_2 g_{bc}-2P^{bcd}\delta_1 \Gamma_{\phantom{e}db}^{e}\delta_2g_{ec}+\delta_1 S^{ab}\delta_2g_{ab}\\
    &+\displaystyle\sum_{i=1}^{m-1}\delta_1 T_{i}^{abcda_1...a_i}\delta_2\nabla_{(a_1}...\nabla_{a_i)} R_{abcd}-(1\leftrightarrow 2)\;,
\end{split}
\label{eq: higherDerivativeSymplecticForm}
\end{equation}
with $P^{bcd}=\epsilon_a P^{abcd}$, $S^{ab}$ and $T_i^{abcd a_1...a_i}$ are functions of the metric, the Riemann tensor and its covariant derivatives. 

To derive the generalized volume $W_{\text{gen}}$, we need to evaluate the symplectic current form with $\delta_{1}g=\delta_{Y}g=\mathcal{L}_{Y}g$ and $\delta_{2}g=\delta g$. Substituting this in and using $\mathcal{L}_{Y}g_{ab}|_{\Sigma}=0$ and $\nabla_{d}(\delta_{Y}g_{ab})|_{\Sigma}=N^{-1}n_{d}h_{ab}$ with $N=1/2\alpha$, we find
\begin{equation}
\begin{split}
    \omega(g,\delta_Y g,\delta g)|_\Sigma=&-2P^{bcd}\delta_Y \Gamma_{\phantom{e}db}^e\delta g_{ec}-2\delta P^{bcd}\nabla_d\delta_Y g_{bc}\\
    =&-P^{bcd}g^{ef}(\nabla_b\delta_Y g_{df}+\nabla_d\delta_Y g_{bf}-\nabla_f\delta_Y g_{db})\delta g_{ec}-2\delta P^{bcd}\nabla_d\delta_Y g_{bc}\\
    =&-\frac{2}{N}\left[2\delta P^{bcd}n_d h_{bc}+P^{bcd}(n_b h_{d}^e+n_dh_{b}^e-n^e h_{db})\delta g_{ec}\right]\;.
\end{split}
\label{eq:sympcurrYgen}\end{equation}
 Here we used that we are deforming away from vacuum AdS, which is maximally symmetric such that the covariant derivative of the Riemann tensor in empty AdS vanishes identically, as does its Lie derivative with respect $Y$ when evaluated on the maximal hypersurface $\Sigma$. Moreover, we used that tensors $P^{abcd},\ S^{ab}$ and $T_i^{abcd a_1...a_i}$ are all functions of the metric and Riemann tensor, such that their Lie derivatives will be zero when evaluated on $\Sigma$. 

 To rewrite the symplectic current (\ref{eq:sympcurrYgen}) as a variation of some scalar function, define \cite{Bueno:2016gnv}
 \beq F^{abcd}\equiv P^{abcd}-P_{0}(g^{ac}g^{bd}-g^{ad}g^{bc})\;,\eeq
 characterizing the difference between $P^{abcd}$ and its background value $P^{abcd}_{\text{MSS}}$. By construction, $F^{abcd}$ vanishes in empty AdS. 
The tensor $F^{abcd}$ is introduced since such that any contribution (\ref{eq:sympcurrYgen}) which includes $\delta F^{abcd}$ may be recast as a total variation of that object because any term multiplied by $F^{abcd}$ will vanish when evaluated about empty AdS. Indeed, consider the first term in the last line of (\ref{eq:sympcurrYgen}):
\begin{equation}
\begin{split}
    2n_dh_{bc}\delta P^{bcd}=&2n_d h_{bc}\delta(\epsilon_a F^{abcd})+2P_0n_dh_{bc}\delta[\epsilon_{a}(g^{ac}g^{bd}-g^{ad}g^{bc})]\\
    =&2\delta(n_d h_{bc}\epsilon_a F^{abcd})+2P_0n_dh_{bc}\delta[\epsilon_{a}(g^{ac}g^{bd}-g^{ad}g^{bc})]\\
    =&2\delta(n_d h_{bc}\epsilon_a P^{abcd})-2P_0\epsilon_{a}(g^{ac}g^{bd}-g^{ad}g^{bc})\delta(n_dh_{bc})\;.
\end{split}
\end{equation}
 To arrive to the second line we used $2n_{d}h_{bc}\delta(\epsilon_{a}F^{abcd})=2\delta(n_{d}h_{bc}\epsilon_{a}F^{abcd})-2F^{abcd}\delta(n_{d}h_{bc}\epsilon_{a})$, subsequently dropping the term proportional to $F^{abcd}$. The final follows from replacing $F^{abcd}$ with $P^{abcd}$. Taking into account that $\epsilon_a=-n_a\wedge \epsilon_{\Sigma}$ (with $\epsilon_{\Sigma}$ being the induced volume form in $\Sigma$), $\delta n_{a}=-\frac{1}{2}n_a n^{b}n^c\delta g_{bc}$ (cf. \cite{Jiang:2018sqj}), and $h_{ab}=g_{ab}+n_an_b$, this becomes
\begin{equation}
    2n_dh_{bc}\delta P^{bcd}=-2\delta(\epsilon_{\Sigma} P^{abcd}n_a n_d h_{bc})-P_0 \epsilon_{\Sigma} n^an^b\delta g_{ab}(d-2)+2P_0 \epsilon_{\Sigma} g^{bc}\delta g_{bc}\;.
\end{equation}
Meanwhile, replacing $P^{abcd}$ for $F^{abcd}$ in the second term in the bottom line of (\ref{eq:sympcurrYgen}) yields
\begin{equation}
    P^{bcd}(n_bh_{d}^e+n_d h_{b}^e-h_{bc}u^e)\delta g_{ec}=\epsilon_a [F^{abcd}+P_0(g^{ac}g^{bd}-g^{ad}g^{bc})](n_bh_{d}^e+n_d h_{b}^e-h_{bc}u^e)\delta g_{ec}.
\end{equation}
Expanding each induced metric, using $P^{abcd}n_a n_b=0$ due to the antisymmetry in the first indices of $P^{abcd}$, and dropping the term proportional to $F^{abcd}$ gives
\begin{equation}
    P^{bcd}(n_bh_{d}^e+n_d h_{b}^e-h_{bc}u^e)\delta g_{ec}=-P_0\epsilon_{\Sigma}(g^{ab}+n^an^b(1-d))\delta g_{ab}\;.
\end{equation}

Putting everything together, the symplectic current (\ref{eq:sympcurrYgen}) is
\begin{equation}
\begin{split}
\omega(g,\delta_Y g,\delta g)=&-\frac{1}{N}\left[-2\delta(P^{abcd}\epsilon_{\Sigma} n_{a}n_d h_{bc})+P_0\epsilon_{\Sigma} n^an^n\delta g_{ab}+P_0\epsilon_{\Sigma} g^{ab}\delta g_{ab}\right]\\
=&\frac{2}{N}\left[\delta(P^{abcd}\epsilon_{\Sigma} n_{a}n_d h_{bc})-\frac{1}{2}P_0\epsilon_{\Sigma} h^{ab}\delta g_{ab}\right].
\end{split}
\end{equation}
Lastly, using  $\delta\epsilon_{\Sigma}=\frac{1}{2}\epsilon_{\Sigma} h^{ab}\delta g_{ab}$, we find
\begin{equation}
\begin{split}
\omega(g,\delta_Y g,\delta g)|_{\Sigma}=\frac{2}{N}\delta\left[\epsilon_{\Sigma}(P^{abcd} n_{a}n_d h_{bc}-P_0)\right]\;,
\end{split}
\end{equation}
as reported in the main text.

\section{ADM formalism} \label{app:ADMforms}

Here we perform the ADM decomposition of a general non-minimally coupled dilaton theory of gravity, of which (semi-) classical JT gravity is a special case. Our approach follows the presentation of \cite{Dyer:2008hb}.

\subsubsection*{Geometric set-up}

To carry out the ADM formulation, foliate the $d$-dimensional Lorentzian spacetime $\tilde{\mathcal{M}}$ with hypersurfaces $\Sigma_{t}$ by introducing a global time function $t(x)$, where $x^{\mu}$ ($\mu=0,1,..,d-1$) are coordinates on $\tilde{\mathcal{M}}$, and $\Sigma_{t}$ are level sets of $t(x)$. Coordinates on hypersurfaces $\Sigma_{t}$ are denoted by $y^{a}(x^{\mu})$ ($a=1,...d-1$). Together, $t$ and $y^{a}$ comprise a new coordinate system on $\tilde{\mathcal{M}}$, with coordinate basis vectors $t^{\mu}=\frac{\partial x^{\mu}}{\partial t}$ and $e^{\mu}_{a}=\frac{\partial x^{\mu}}{\partial y^{a}}$. The bulk metric on $\tilde{\mathcal{M}}$ is denoted by $g_{\mu\nu}$ while the induced metric on $\Sigma_{t}$ is $h_{\mu\nu}=g_{\mu\nu}+n_{\mu}n_{\nu}$, with $n^{\mu}$ being the future-directed timelike unit normal to $\Sigma_{t}$ and orthogonal to $e^{\mu}_{a}$. Projecting onto the hypersurface, the induced metric on $\Sigma_{t}$ is $h_{ab}=e^{\mu}_{a}e^{\nu}_{b}h_{\mu\nu}$, and the extrinsic curvature is $K_{ab}=e^{\mu}_{a}e^{\nu}_{b}\nabla_{\mu}n_{\nu}=\frac{1}{2}e^{\mu}_{a}e^{\nu}_{b}\mathcal{L}_{n}g_{\mu\nu}$. Meanwhile the trace of the extrinsic curvature is $K=h^{ab}K_{ab}=\nabla_{\mu}n^{\mu}$.

The timelike boundary $\mathcal{B}$ of the bulk manifold has a radially outward pointing normal vector $r^{\mu}$, and is described by coordinates $z^{i}$ ($i=0,2,...,d-1$) with an induced metric $\gamma_{\mu\nu}=g_{\mu\nu}-r_{\mu}r_{\nu}$. We will insist that surfaces $\Sigma_{t}$ intersect $B$ orthogonally such that $g_{\mu\nu}r^{\mu}n^{\nu}=0$ on $B$, implying $r^{\mu}$ is parallel to $\Sigma_{t}$ on $B$, \emph{i.e.}, $r^{\mu}=r^{a}e^{\mu}_{a}$. Intersections of $\Sigma_{t}$ with $B$ are denoted by $S_{t}=\Sigma_{t}\cap B$. The $(d-2)$-dimensional slices $S_{t}$ form a spacelike foliation of $B$ and denote coordinates on $S_{t}$ by $\theta^{A}$ ($A=2,3,...d-1$), and $\sigma_{AB}$ as the induced metric on $S_{t}$. The extrinsic curvatures of $\mathcal{B}$ and of $S_{t}$ are, respectively, $\mathcal{K}_{ij}$ and  $k_{AB}$. 

In ADM variables $(t,y^{a})$, the bulk $d$-dimensional spacetime line element is 
\beq ds^{2}=g_{\mu\nu}dx^{\mu}dx^{\nu}=-N^{2}dt^{2}+h_{ab}(N^{a}dt+dy^{a})(N^{b}dt+dy^{b})\;. \label{eq:ADMmet}\eeq
where one uses $dx^{\mu}(t,y^{a})=t^{\mu}dt+e^{\mu}_{a}dy^{a}$. Here $N$ and $N^{a}$ are the lapse and shift, respectively, and decompose the global time function vector defining the foliation as $t^{\mu}=Nn^{\mu}+N^{a}e^{\mu}_{a}$. Moreover, in the $(t,y^{a})$ coordinate system, the normal $n_{\mu}=-N\partial_{\mu}t$ obeys
\beq n_{0}=-N\;,\quad n_{a}=0\;,\quad n^{0}=\frac{1}{N}\;,\quad n^{a}=-\frac{N^{a}}{N}\;.\label{eq:normalinADM}\eeq
Additionally, in ADM variables we have $\sqrt{|g|}=N\sqrt{|h|}$, while the extrinsic curvature $K_{ab}$ is 
\beq K_{ab}=\frac{1}{2N}(\dot{h}_{ab}-\nabla_{a}N_{b}-\nabla_{b}N_{a})\;,\label{eq:KabADM}\eeq
and the $d$-dimensional Ricci scalar $R$ decomposes as
\beq R=\bar{R}-(K^{2}-K_{ab}K^{ab})-2\nabla_{\alpha}(n^{\beta}\nabla_{\beta}n^{\alpha}-n^{\alpha}K)\;,\label{eq:RiccidecompADM}\eeq
where $\bar{R}$ is the $(d-1)$-dimensional Ricci scalar constructed from the induced metric $h_{ab}$.

\subsubsection*{ADM Lagrangian for general theories}

Consider a non-minimally coupled dilaton theory of gravity in $d$ dimensions, characterized by
\beq I=I_{\text{EH}}+I_{\phi}+I_{\text{GHY}}\;,\eeq
with
\beq 
\begin{split}
&I_{\text{EH}}=\int _{\tilde{\mathcal{M}}}d^{d}x\sqrt{-g}f(\phi)R\;,\\
&I_{\phi}=\int_{\tilde{\mathcal{M}}}d^{d}x\sqrt{-g}\left[-\frac{1}{2}\lambda(\phi)g^{\alpha\beta}\partial_{\alpha}\phi\partial_{\beta}\phi-U(\phi)\right]\;,\\
&I_{\text{GHY}}=2\int_{\partial\tilde{\mathcal{M}}}d^{d-1}x\sqrt{|h|}Kf(\phi)\;.
\end{split}
\label{eq:scalartenact1}\eeq
Substituting in the decomposition of the Ricci scalar (\ref{eq:RiccidecompADM}), the Einstein-Hilbert contribution (\ref{eq:scalartenact1}) is 
\beq I_{\text{EH}}=\int_{\tilde{\mathcal{M}}}d^{d}x\sqrt{-g}f[\bar{R}-K^{2}+K_{ab}^{2}-2\nabla_{\alpha}(n^{\beta}\nabla_{\beta}n^{\alpha}-n^{\alpha}K)]\;.\label{eq:EHact1}\eeq
The GHY boundary action, meanwhile, splits into three terms coming from the three boundary contributions comprising $\partial \tilde{\mathcal{M}}$:
\beq I_{\text{GHY}}=2\oint_{\mathcal{B}}d^{d-1}z\sqrt{-\gamma}\mathcal{K}f+2\oint_{\Sigma_{1}}d^{d-1}g\sqrt{h}Kf-2\oint_{\Sigma_{2}}d^{d-1}y\sqrt{h}Kf\;,\label{eq:GHYv1}\eeq
with $\mathcal{K}$ being the trace of the extrinsic curvature on $\mathcal{B}$, $\gamma^{ij}\mathcal{K}_{ij}=\mathcal{K}$.\footnote{Notice the integral over $\Sigma_{2}$ has an additional minus sign since, by our convention, the normal is directed inward for spacelike surfaces, whereas for $\Sigma_{2}$ it is directed outward.}
Performing integration by parts on the last term in the Einstein-Hilbert action,\footnote{We also use $\nabla_{\beta}n^{2}=0\Rightarrow n^{\alpha}\nabla_{\beta}n_{\alpha}=0$, and $r_{\alpha}n^{\alpha}=0$.} one finds a cancellation of the $\Sigma_{1,2}$ contributions to the GHY term, leaving, after some additional simplifications \cite{Dyer:2008hb}
\beq
\begin{split}
 \hspace{-3mm} I_{\text{EH}}+I_{\text{GHY}}&=\int_{\tilde{\mathcal{M}}}d^{d}x\sqrt{h}\left[Nf(\bar{R}-K^{2}+K_{ab}^{2})+2f'(h^{ab}(\partial_{a}N)\partial_{b}\phi-K\dot{\phi}+KN^{a}\partial_{a}\phi)\right]\\
&+2\oint_{\mathcal{B}}d^{d-1}z\sqrt{\sigma}Nfk\;.
\end{split}
\label{eq:IGIGHYfinal}\eeq
where $\sqrt{-g}=N\sqrt{h}$, and $f'(\phi)=\frac{df}{d\phi}$.
 Meanwhile, the scalar field action $I_{\phi}$ (\ref{eq:scalartenact1}) in ADM variables is
\beq 
\begin{split}
I_{\phi}&=\int_{\tilde{\mathcal{M}}}d^{d}x\left(\frac{\lambda\sqrt{h}}{2N}\left[\dot{\phi}(\dot{\phi}-2N^{a}\partial_{a}\phi)-N^{2}h^{ab}(\partial_{a}\phi)(\partial_{b}\phi)+(N^{a}\partial_{a}\phi)^{2}\right]-N\sqrt{h}U\right)\;.
\end{split}
\label{eq:scalardilatonact}\eeq

Collectively, we assemble the actions (\ref{eq:IGIGHYfinal}) and (\ref{eq:scalardilatonact}) into a single ADM action $I_{\text{ADM}}$
\beq I_{\text{ADM}}=\int dt L_{\text{ADM}}+2\oint_{\mathcal{B}}d^{d-1}z\sqrt{\sigma}Nfk\;,\label{eq:ADMactioncompact}\eeq
where $L_{\text{ADM}}$ is the ADM Lagrangian:
\beq L_{\text{ADM}}=\int_{\Sigma_{t}}d^{d-1}y\mathcal{L}_{\text{ADM}}\;,\eeq
with ADM Lagrangian density $\mathcal{L}_{\text{ADM}}$
\beq 
\begin{split}
&\mathcal{L}_{\text{ADM}}=\sqrt{h}\biggr\{Nf[\bar{R}-K^{2}+K_{ab}^{2}]+2f'h^{ab}(\partial_{a}N)(\partial_{b}\phi)\\
&+\frac{\lambda}{2N}\left[\dot{\phi}(\dot{\phi}-2N^{a}\partial_{a}\phi)-N^{2}h^{ab}(\partial_{a}\phi)(\partial_{b}\phi)+(N^{a}\partial_{a}\phi)^{2}\right]-NU(\phi)-2f'K(\dot{\phi}-N^{a}\partial_{a}\phi)\biggr\}\;.
\end{split}
\label{eq:ADMLagrangiandensity}\eeq
Since in ADM variables the extrinsic curvature $K_{ab}$ decomposes as (\ref{eq:KabADM})
the ADM Lagrangian is a function of the dynamical variables $\{h_{ab},\dot{h}_{ab},N,N^{a},\phi,\dot{\phi}\}$. Notice in the limit $f\to1$ and there is no $\phi$ dependence, the Lagrangian (\ref{eq:ADMLagrangiandensity}) reduces to the usual ADM Lagrangian of general relativity (plus the boundary surface term).

\subsubsection*{ADM Hamiltonian for general theories}

To transition to the Hamiltonian formulation, let us first determine the canonical momenta conjugate the dynamical variables in the ADM action. Since there are now time derivatives of the lapse $N$ or shift $N^{a}$, we immediately find
\beq \pi_{N}\equiv\frac{\delta L_{\text{ADM}}}{\delta \dot{N}}=0\;,\quad \pi^{a}_{N^{a}}\equiv\frac{\delta L_{\text{ADM}}}{\delta\dot{N}^{a}}=0\;.\eeq
We thus obtain the primary constraints of the theory, $\pi_{N}\approx0$ and $\pi_{a}^{N^{a}}\approx0$. The momenta conjugate to $\phi$ and $h_{ab}$ are, respectively,
\beq \pi_{\phi}=\frac{\delta L_{\text{ADM}}}{\delta\dot{\phi}}=\frac{\sqrt{h}\lambda}{N}\left(\dot{\phi}-N^{a}\partial_{a}\phi\right)-2\sqrt{h}f'K\;,
\label{eq:piphi}\eeq
\beq
\begin{split}
\pi^{ab}&=\frac{\delta L_{\text{ADM}}}{\delta\dot{h}_{ab}}=\sqrt{h}\left[f(K^{ab}-Kh^{ab})-\frac{f'}{N}h^{ab}(\dot{\phi}-N^{c}\partial_{c}\phi)\right]\;,
\end{split}
\label{eq:conjpiab}\eeq
where we used $K_{ab}$ in ADM variables (\ref{eq:KabADM}) such that $\delta K_{ab}=\frac{1}{2N}\delta\dot{h}_{ab}$.
Importantly, we may invert the conjugate momenta expressions, allowing us to replace $\dot{\phi}$ and $K_{ab}$ (and $K$) in terms of the canonical (phase space) variables:
\beq \dot{\phi}=\frac{N}{\sqrt{h}}\left(\frac{(d-2)f\pi_{\phi}-2f'\pi}{2(d-1)(f')^{2}+(d-2)f\lambda}\right)+N^{a}\partial_{a}\phi\;,\eeq
\beq K_{ab}=\frac{1}{f}\frac{\pi_{ab}}{\sqrt{h}}-\frac{h_{ab}}{\sqrt{h}}\left(\frac{\pi_{\phi}f'+2\pi\frac{(f')^{2}}{f}+\pi\lambda}{2(d-1)(f')^{2}+(d-2)f\lambda}\right)\;,\eeq
with $\pi\equiv h^{ab}\pi_{ab}$. To show this, it is useful, first take the trace of $\pi^{ab}$ to solve for $K$, 
\beq K=-\frac{1}{\sqrt{h}}\left(\frac{(d-1)f'\pi_{\phi}+\lambda \pi}{(d-2)\lambda f+2(d-1)(f')^{2}}\right)\;,\label{eq:extKmom}\eeq
and substitute this into $\pi_{\phi}$ to solve for $\dot{\phi}$.\footnote{Note there is a typo in $K_{ab}$ in Eq. (6.18) of \cite{Dyer:2008hb}.}

Note that the map from velocities $(\dot{\phi},\dot{h}_{ab})$ to momenta $(\pi_{\phi},\pi_{ab})$ is nonsingular even when the scalar kinetic term vanishes $\lambda\to0$. The case $f\to\text{const}\neq0$ is also well behaved, provided $\lambda\neq0$, as the scalar has dynamics stemming from the kinetic term. The case $f\to\text{const}$ and $\lambda\to0$ is singular, however, this is because now the scalar field loses its dynamics \cite{Dyer:2008hb}. Also note the solution here is non-singular in the limit $d\to2$. 

The ADM Hamiltonian is
\beq H_{\text{ADM}}=\int_{\Sigma_{t}}d^{d-1}y\mathcal{H}_{\text{ADM}}+H_{B}\;,\eeq
 where $\mathcal{H}_{\text{ADM}}$ is the Hamiltonian density given by,
\beq \mathcal{H}_{ADM}=\pi^{ab}\dot{h}_{ab}+\pi_{\phi}\dot{\phi}+\lambda_{N} \pi_{N}+\lambda^{a}\pi_{a}^{N^{a}}-\mathcal{L}_{\text{ADM}}\;,\label{eq:Hamdensgen}\eeq
and $H_{B}$ is a boundary Hamiltonian associated with the boundary term in the ADM action (\ref{eq:ADMactioncompact}). Here we have introduced Lagrange multipliers $\lambda_{N}$ and $\lambda^{a}$, as is standard for constrained Hamiltonian systems. After some algebra, the total ADM Hamiltonian is\footnote{There appears to be a minor discrepancy between our expression and Eq. (6.22) in \cite{Dyer:2008hb}, however, we agree with the combination of their Eqs. (6.19) -- (6.21).} 
\beq
\begin{split}
&H_{\text{ADM}}=\int_{\Sigma_{t}}d^{d-1}y\sqrt{h}\biggr\{N\biggr[\frac{2\pi^{ab}}{\sqrt{h}}K_{ab}-f[\bar{R}-K^{2}+K_{ab}^{2}]+2\nabla^{a}(f'\nabla_{a}\phi)\\
&+\frac{\pi^{2}_{\phi}}{2\lambda h}+\frac{2f'K}{\lambda}\frac{\pi_{\phi}}{\sqrt{h}}+2(f')^{2}\frac{K^{2}}{\lambda}+\frac{\lambda}{2}h^{ab}(\partial_{a}\phi)(\partial_{b}\phi)+U\biggr]+N_{a}\left(\frac{\pi_{\phi}}{\sqrt{h}}\partial^{a}\phi-2\nabla_{b}\left(\frac{\pi^{ab}}{\sqrt{h}}\right)\right)\biggr\}\\
&+2\oint_{S_{t}}d^{d-2}\theta\sqrt{\sigma}\left[r_{a}N_{b}\frac{\pi^{ab}}{\sqrt{h}}-N(fk+r^{a}f'\partial_{a}\phi)\right]\;.
\end{split}
\eeq
More compactly, we may introduce the super-Hamiltonian $\mathcal{H}$ and super-momentum $\mathcal{H}^{a}$ as terms proportional to the Lagrange multipliers $N$ and $N_{a}$ in the integral over $\Sigma_{t}$. That is, 
\beq 
\begin{split}
\mathcal{H}&\equiv\biggr[\frac{2\pi^{ab}}{\sqrt{h}}K_{ab}-f[\bar{R}-K^{2}+K_{ab}^{2}]\\
&+2\nabla^{a}(f'\nabla_{a}\phi)+\frac{\pi^{2}_{\phi}}{2\lambda h}+\frac{2f'K}{\lambda}\frac{\pi_{\phi}}{\sqrt{h}}+2(f')^{2}\frac{K^{2}}{\lambda}+\frac{\lambda}{2}h^{ab}(\partial_{a}\phi)(\partial_{b}\phi)+U\biggr]\;,
\end{split}
\eeq
\beq \mathcal{H}^{a}\equiv \left(\frac{\pi_{\phi}}{\sqrt{h}}\partial^{a}\phi-2\nabla_{b}\left(\frac{\pi^{ab}}{\sqrt{h}}\right)\right)\;.\eeq
Using this, the ADM Hamiltonian $H_{\text{ADM}}$ is:
\beq H_{\text{ADM}}=\int_{\Sigma_{t}}d^{d-1}y\sqrt{h}(N\mathcal{H}+N_{a}\mathcal{H}^{a})+\text{terms on}\;\; S_{t}\;.
\eeq
From this, it is easy to read off the Hamiltonian density as $\mathcal{H}_{\text{ADM}}=\sqrt{h}(N\mathcal{H}+N_{a}\mathcal{H}^{a})$.

\subsection*{Special limits}

Let us consider a few relevant special limits. 

\vspace{2mm}

\noindent \textbf{No kinetic term.} First consider the case when $\lambda=0$. The ADM Lagrangian (\ref{eq:ADMLagrangiandensity}) is
\beq \mathcal{L}_{\text{ADM}}=\sqrt{h}\biggr\{Nf[\bar{R}-K^{2}+K_{ab}^{2}]+2f'h^{ab}(\partial_{a}N)(\partial_{b}\phi)-NU(\phi)-2f'K(\dot{\phi}-N^{a}\partial_{a}\phi)\biggr\}\;.\eeq
The conjugate momentum (\ref{eq:piphi}) simplifies to
\beq \pi_{\phi}=-2\sqrt{h}f' K\;,\eeq
while $\pi^{ab}$ (\ref{eq:conjpiab}) remains the same.
The inverse relations become
\beq \dot{\phi}=\frac{N}{\sqrt{h}}\left(\frac{(d-2)f\pi_{\phi}-2f'\pi}{2(d-1)(f')^{2}}\right)+N^{a}\partial_{a}\phi\;,\eeq
\beq K_{ab}=\frac{1}{f}\frac{\pi_{ab}}{\sqrt{h}}-\frac{h_{ab}}{\sqrt{h}}\left(\frac{\pi_{\phi}f'+2\pi\frac{(f')^{2}}{f}}{2(d-1)(f')^{2}}\right)\;.\eeq
The constraints are
\beq 
\begin{split}
\mathcal{H}&\equiv\biggr[\frac{2\pi^{ab}}{\sqrt{h}}K_{ab}-f[\bar{R}-K^{2}+K_{ab}^{2}]+2\nabla^{a}(f'\nabla_{a}\phi)+U\biggr]\;,
\end{split}
\eeq
\beq \mathcal{H}^{a}\equiv \left(\frac{\pi_{\phi}}{\sqrt{h}}\partial^{a}\phi-2\nabla_{b}\left(\frac{\pi^{ab}}{\sqrt{h}}\right)\right)\;.\eeq

\vspace{2mm}

\noindent \textbf{General relativity.} To recover general relativity with a cosmological constant, in addition to $\lambda=0$, we set $f=1$ and $U(\phi)=2\Lambda$. Then, the ADM Lagrangian is
\beq \mathcal{L}_{\text{ADM}}=N\sqrt{h}[\bar{R}-2\Lambda-K^{2}+K_{ab}^{2}]\;.\eeq
 Further, the canonical momenta simplify to $\pi_{\phi}=0$, and 
\beq \pi^{ab}=\sqrt{h}(K^{ab}-Kh^{ab})\;,\quad K_{ab}=\frac{1}{\sqrt{h}}\pi_{ab}-\frac{1}{\sqrt{h}}\frac{h_{ab}}{(d-2)}\pi\;,\eeq
such that the constraints are
\beq 
\mathcal{H}\equiv\biggr[\frac{2\pi^{ab}}{\sqrt{h}}K_{ab}-[\bar{R}-2\Lambda-K^{2}+K_{ab}^{2}]\biggr]\;,\quad  \mathcal{H}^{a}\equiv -2\nabla_{b}\left(\frac{\pi^{ab}}{\sqrt{h}}\right)\;.\eeq

\vspace{2mm}

\noindent \textbf{2D dilaton gravity.} In two-dimensions $K_{ab}=Kh_{ab}$. Hence, the ADM Lagrangian is 
\beq 
\begin{split}
&\mathcal{L}_{\text{ADM}}=\sqrt{h}\biggr\{Nf\bar{R}+2f'h^{ab}(\partial_{a}N)(\partial_{b}\phi)-NU(\phi)-2f'K(\dot{\phi}-N^{a}\partial_{a}\phi)\\
&+\frac{\lambda}{2N}\left[\dot{\phi}(\dot{\phi}-2N^{a}\partial_{a}\phi)-N^{2}h^{ab}(\partial_{a}\phi)(\partial_{b}\phi)+(N^{a}\partial_{a}\phi)^{2}\right]\biggr\}\;.
\end{split}
\label{eq:ADMLagrangiandensity2D}\eeq
The  conjugate momenta are
\beq \pi^{ab}=\pi h^{ab}\;,\quad \pi=-\sqrt{h}\frac{f'}{N}(\dot{\phi}-N^{c}\partial_{c}\phi)\;,\quad \pi_{\phi}=-\frac{\lambda\pi}{f'}-2\sqrt{h}f' K\;,\label{eq:conjmomgenapp}\eeq
leading to the following constraints
\beq 
\begin{split}
\hspace{-2mm}\mathcal{H}&\equiv\biggr[\frac{2\pi}{\sqrt{h}}K-f\bar{R}+2\nabla^{a}(f'\nabla_{a}\phi)+\frac{\pi^{2}_{\phi}}{2\lambda h}+\frac{2f'K}{\lambda}\frac{\pi_{\phi}}{\sqrt{h}}+2(f')^{2}\frac{K^{2}}{\lambda}+\frac{\lambda}{2}h^{ab}(\partial_{a}\phi)(\partial_{b}\phi)+U\biggr]\;,
\end{split}
\eeq
\beq \mathcal{H}^{a}\equiv \left(\frac{\pi_{\phi}}{\sqrt{h}}\partial^{a}\phi-2\nabla_{b}\left(\frac{\pi^{ab}}{\sqrt{h}}\right)\right)\;.\eeq
Note that classical JT gravity corresponds to $U=2\Phi\Lambda$, $f=(\Phi_{0}+\Phi)$, and $\lambda=0$, while the semi-classical Polyakov correction has $f=\chi$, $\lambda=-2$ and $U=0$.\footnote{Here we neglect the counterterm because it only shifts the boundary contribution to the Hamiltonian and does not alter the constraints.}

\bibliographystyle{JHEP}
\bibliography{refs-IAMFL}

\end{document}